\documentclass[11pt,letterpaper]{article}

\usepackage{url}
\usepackage{fullpage}

\usepackage[subtle,mathspacing=normal]{savetrees}
\usepackage{amsmath}
\usepackage{algorithm}
\usepackage[noend]{algpseudocode}
\usepackage{amssymb}
\usepackage{amsthm}
\usepackage{color}
\usepackage{array}
\usepackage{xy}
\usepackage{setspace}
\usepackage{varwidth}
\usepackage{multicol}
\usepackage{algorithm}
\usepackage{hyperref}
\usepackage{thm-restate}
\usepackage{todonotes}

\newcommand{\poly}{\operatorname{poly}}

 \newcommand{\E}{\mathbb{E}}

\newtheorem{thm}{Theorem}[section]

\theoremstyle{remark}

\newtheorem{theorem}{Theorem}[section]

\newtheorem{lemma}[thm]{Lemma}
\newtheorem{proposition}[thm]{Proposition}
\newtheorem{claim}[thm]{Claim}
\newtheorem{corollary}[thm]{Corollary}

\theoremstyle{remark}

\newcommand{\defn}[1]{\textbf{\emph{#1}}}
\renewcommand{\paragraph}[1]{\vspace{.2 cm} \noindent \textbf{#1}}

\usepackage{authblk}

\renewcommand{\epsilon}{\varepsilon}
\renewcommand{\emptyset}{\varnothing}
\newcommand{\mc}[1]{\mathcal{#1}}
\newcommand{\card}[1]{|#1|}

\newcommand{\vol}{\mathsf{vol}}

\newcommand{\R}{\mc{R}}
\begin{document}
\title{The Power of Two-Choice Linear Probing}
\date{}
\author[1]{Amir Azarmehr\thanks{Supported in part by NSF CAREER grant CCF-2442812 and a Google Research Award.}}
\author[2]{Michael A. Bender\thanks{Supported in part by NSF grants CCF-2247577 and IIS-2551616.}}
\author[3]{William Kuszmaul\thanks{Supported in part by NSF grant CNS-2504471, NSF CAREER grant CCF-2542165, and a Jane Street grant.}}
\author[3]{Rose Silver\thanks{Supported in part by the Carnegie Mellon University CyLab Presidential Fellowship.}}
\affil[1]{Northeastern University}
\affil[2]{Stony Brook University}
\affil[3]{Carnegie Mellon University}
\maketitle

\begin{abstract}
This paper considers the following basic question: If an (ordered) linear-probing hash table is allowed \emph{two} hash functions, instead of one, how does this change the expected insertion and query time, as a function of the load factor $1 - \epsilon$? We prove that the \emph{greedy two-choice insertion strategy} achieves polynomially better bounds than the single choice algorithm, but that one can even do \emph{much better} by using more sophisticated non-greedy strategies. Specifically, we show that there is an insertion strategy that does not evict elements (once an element is inserted, its hash choice is fixed) and that achieves expected query time $O(\log \epsilon^{-1})$ with expected insertion time $O(\epsilon^{-1})$. We then further show that, if one is allowed to evict elements (i.e., to change over time which hash function a given element uses), then it is possible to achieve expected query time $O(1)$ with expected insertion time $O(\epsilon^{-1/2})$. This final result achieves an expected query time of $O(1)$ even when the hash table is filled to $100\%$ full. Combined, the results reveal that there is a surprisingly strong ``power of two choices'' phenomenon for linear-probing hash tables, allowing for a two-choice hash table to achieve significantly better bounds than what might at first seem to be possible.
\end{abstract}

\section{Introduction}

The \defn{linear probing hash table} is one of the simplest and most widely used data structures in computer science. It stores elements in an array $A$ of some size $n$, and supports insertions $\texttt{Insert}(x)$ by placing $x$ in the first free slot out of positions $h(x), h(x) + 1, \ldots \pmod{n}$, where $h(x) \in [n]$ is a uniformly random hash. Early work on linear probing \cite{Peterson57} conjectured that, when the hash table is filled to load factor $1 - \epsilon$ (meaning we have performed $(1 - \epsilon)n$ insertions), the next insertion should take expected time $\Theta(\epsilon^{-1})$. However, in 1962, Knuth showed this conjecture to be false, establishing instead a bound of $\Theta(\epsilon^{-2})$ \cite{Knuth63}. This quadratic blowup in insertion time is due to a \emph{clustering effect} in which elements tend to group together into long \emph{runs} of occupied slots.

Clustering also hurts query time. As Peterson pointed out in 1957 \cite{Peterson57}, if the insertion time \emph{were} $\Theta(\epsilon^{-1})$, then the expected time to query a \emph{random} element out of those present would be $\Theta(\int_{\delta = \epsilon}^1 \delta^{-1} \, d\delta) = \Theta(\log \epsilon^{-1})$. Unfortunately, due to Knuth's clustering effect, the expected query time is instead $\Theta(\int_{\delta = \epsilon}^1 \delta^{-2} \, d\delta) = \Theta(\epsilon^{-1})$.
Thus, whereas clustering causes a \emph{quadratic} blowup in insertion time, it causes an \emph{exponential} blowup in query time.

Note that, although the above query bounds concern a \emph{random element out of those present}, there is a standard modification to linear probing that extends the bound to hold for an arbitrary query: Just maintain the invariant that, within each run of occupied slots, the elements are in sorted order of hash. This variation, known as \emph{ordered linear probing}, was first introduced by Amble and Knuth in 1974 \cite{AmbleKn74}, and guarantees that every query has the same expected time of $\Theta(\epsilon^{-1})$. We will use variations of ordered linear probing throughout this paper.

Knuth's result motivates a natural question: Can linear probing be modified to avoid the effects of clustering? For \emph{insertions}, the answer is yes: recent work by Bender et al.~\cite{bender2022linear} constructs a version of linear probing, which they call \emph{graveyard hashing}, that achieves expected time $\Theta(\epsilon^{-1})$ for insertions (and queries). But, for \emph{queries}, the answer is no: Even in the offline setting, no matter how we choose to order the elements in the hash table, the expected number of probes to find a random element remains the same \cite{Peterson57} (or larger) as if we just used classical linear probing.

A natural way to get around this bottleneck is to allow the hash table to use \emph{two hash functions} $h_1(x), h_2(x)$, and to place each element $x$ using linear probing with one of the two hash functions (see, e.g., \cite{nijalingappavertical,gurumurthy2018simd}). By strategically choosing which hash function to use, one could hope to either reduce (or maybe even completely eliminate) the effects of clustering not just for insertions, but also for queries.\footnote{This may feel, at first glance, similar to (bucketized) cuckoo hashing \cite{pagh2004cuckoo,dietzfelbinger2007balanced}, in which elements $x$ hash to \emph{bins} $h_1(x), h_2(x)$ (typically of size $\Theta(\log \epsilon^{-1})$), and in which insertions perform ``eviction chains'' to move elements around and find a free slot. However, as discussed in Section \ref{sec:related-work}, although bucketized cuckoo hashing is another example of a two-choice hash table, the two designs lead to remarkably different tradeoff curves.
}

In this paper, we are interested in the following basic questions: Does two-choice linear probing allow for better bounds than classical linear probing? And if so, what is the optimal way to choose between the two hash functions? Despite the simplicity of these questions, they do not appear to have been studied in the theory literature until now. And, as we will see, the questions come with somewhat surprising answers. Two-choice linear probing, when implemented with the correct insertion algorithm, is a \emph{remarkably} powerful way to reduce clustering. It allows us not just to realize the $O(\log \epsilon^{-1})$ query bound envisioned by Peterson \cite{Peterson57}, but to (in our final result) even reduce that query time to $O(1)$, regardless of $\epsilon$.

\paragraph{Analyzing greedy: query time $\epsilon^{-(1 - \Omega(1))}$.} We begin by considering an even more basic question: How does two-choice linear-probing behave if insertions are performed using the simple \defn{greedy strategy}. This strategy places the element $x$ using whichever hash $h_1$ or $h_2$ results in a faster insertion. The greedy strategy is arguably the most natural way to implement two-choice linear probing, and has been employed in several experimental works \cite{nijalingappavertical,gurumurthy2018simd}.

We show that greedy two-choice linear probing achieves polynomially better bounds, for both insertions \emph{and} queries, than classical linear probing:
\begin{restatable}[Greedy two-choice linear probing]{theorem}{thmGreedy}
    Let $\epsilon^{-1} = n^{o(1)}$. There exists a constant $c \in (0, 1)$ such that, at load factor $1 - \epsilon$, greedy two-choice linear probing achieves an expected insertion time of $O(\epsilon^{-(1 + c)})$ and an expected query time of $O(\epsilon^{-c})$.
    \label{thm:greedy}
\end{restatable}

Notice that the greedy algorithm is an inherently \emph{history-dependent} process: the set $S$ of slots that are occupied at any given moment is a function not just of which elements $X$ are present, but also of the \emph{order} in which the elements $X$ were inserted. This comes in stark contrast with classical linear probing, where $S$ is completely determined by $X$. As a consequence, the classical analyses of linear probing \cite{Knuth63,KonheimWe66,DBLP:journals/jacm/MendelsonY80,PaghConstant,bender2022linear} do not seem to extend (at all) to the two-choice setting.

To get around this bottleneck, we first introduce an intermediate result -- a new analysis of \emph{classical} linear probing that allows for us to reason about the behavior of the hash table as a time-evolving process (Section \ref{sec:classic}). The argument comes with the interesting feature that it explicitly illustrates a claim made by Knuth in his original paper, that clustering is fundamentally due to two effects: (1) a positive feedback loop in which larger runs are more likely to receive future insertions, and (2) a globbing effect, in which insertions \emph{combine} together runs that were each already potentially long. This explanation, although intuitively true, is not part of the classical analyses \cite{Knuth63,KonheimWe66,DBLP:journals/jacm/MendelsonY80,PaghConstant,bender2022linear}, which analyze the hash-table in a completely time-independent manner.\footnote{And, indeed, the fact that the analyses are time-independent has led a number of sources \cite{CormenLeRi09, DrozdekSi95, Kruse84, LewisDe91, Morris68, TremblaySo84, WikipediaPrimaryClustering, Smith04} to attribute clustering to only one of the two effects, namely the positive feedback loop, even though, as pointed out by Bender et al.~\cite{bender2022linear}, this effect alone is not enough to change the asymptotics of the hash table.} As such, we suspect that the new analysis, even for standard linear probing, is of independent interest.

We are then able to extend our time-dependent analysis to reason about the behavior of greedy two-choice linear probing. This extension, which requires a number of additional and quite technical arguments, is how we obtain Theorem \ref{thm:greedy}. It remains open what the precise value of $c$ is in Theorem \ref{thm:greedy}. We conjecture based on preliminary experimental work that $c > 0.5$. This suggests that greedy two-choice linear probing reduces but does \emph{not} eliminate clustering -- it still falls short of the ideal bound envisioned by Peterson \cite{Peterson57}.

\paragraph{A non-greedy strategy with $O(\log \epsilon^{-1})$ queries.} Our second result is a \emph{non-greedy} two-choice linear probing algorithm that performs better:

\begin{restatable}[Non-greedy two-choice linear probing]{theorem}{thmNonGreedy}
    Let $\epsilon^{-1} = n^{o(1)}$. There is an implementation of two-choice linear probing that, at load factor $1 - \epsilon$, supports $O(\log \epsilon^{-1})$ expected-time queries and $O(\epsilon^{-1})$ amortized expected-time insertions.
    \label{thm:non-greedy}
\end{restatable}

The insertion algorithm that we use to prove Theorem \ref{thm:non-greedy} is non-greedy in the sense that it will not always use the hash function $h \in \{h_1, h_2\}$ that makes the insertion cheapest. Instead, it selects the hash function in a way that (approximately) achieves what we call the \defn{uniform-free-slot property}: the free slot that the insertion uses is \emph{uniformly random} out of all free slots. Of course, actually achieving this property could be potentially expensive (requiring the insertion to use some very far-away free slot). The main contribution of Theorem \ref{thm:non-greedy} is that there is, in fact, a simple insertion strategy that achieves the uniform-free-slot property without compromising insertion efficiency.

We also prove a matching lower bound: Any insertion algorithm that treats the choice between $h_1$ and $h_2$ as \emph{immutable} (i.e., future insertions cannot change which hash function previous insertions use) must incur an expected insertion time of $\Omega(\epsilon^{-1})$ and an expected query time of $\Omega(\log \epsilon^{-1})$. This makes Theorem \ref{thm:non-greedy} optimal, up to constant factors, out of all immutable two-choice linear probing algorithms. The lower bound is proven in Section \ref{sec:matchinglowerbound}.

\paragraph{Reducing query time to $O(1)$, for \emph{any} $\epsilon$: a non-greedy algorithm with evictions.} Finally, we consider what happens if we allow for an insertion algorithm that not only gets to choose which hash function $h \in \{h_1, h_2\}$ to use for the current insertion, but that also gets to \emph{toggle} which hash function $h \in \{h_1, h_2\}$ other elements $y$ are using.

Here, we discover a quite remarkable phenomenon: it is possible to construct an insertion strategy that reduces the expected query time all the way to $O(1)$, \emph{regardless of the parameter $\epsilon$.} This means that, even when the hash table is filled to $100\%$ full, the expected query time is still $O(1)$.\footnote{Note that Theorem \ref{thm:non-greedyevictions}, unlike the previous theorems in the section, does not impose any restriction on the load factor $\epsilon$.}

\begin{restatable}[Non-greedy two-choice linear probing with evictions]{theorem}{thmEvictions}
    There exists a two-choice linear probing algorithm (with evictions) that, at load factor $1 - \epsilon$, achieves an expected insertion time of $O(\epsilon^{-1/2})$ and an expected query time of $O(1)$.
    \label{thm:non-greedyevictions}
\end{restatable}

We remark that, even in the offline setting, Theorem \ref{thm:non-greedyevictions} comes as somewhat of a surprise. It says that, given $n$ balls with random hashes $h_1, h_2$, it is possible to place them into $n$ slots such that the average distance from each ball $x$'s position to its nearest hash $h_i(x)$ is $O(1)$. In contrast, if one uses a single hash function, then the average distance is necessarily $\Omega(\sqrt{n})$.

Note that Theorem \ref{thm:non-greedyevictions} also comes with (what is basically an unintended) side-effect for insertions -- the expected insertion time drops to $O(\epsilon^{-1/2})$. Thus, every operation takes expected time $o(\epsilon^{-1})$ (a property that, until recently, was not known to be possible \cite{bender2024tight} in any open-addressed hash table).

We remark that the basic idea of rearranging elements in a two-choice linear-probing hash table in order to minimize the query time does not seem to have been studied in \emph{either} the theory literature or the practical literature, until now. Preliminary experiments suggest that this idea is quite promising in practice, allowing for the construction of real-world hash tables that support arbitrarily high load factors while supporting queries in $O(1)$ time with good real-world constants. For example, in experiments on a hash table of size 100,000, we found that simple heuristic-based insertion policies were able to support insertions up to 100\% full, while keeping the average query time down to \emph{just $\approx 2.2$ probes per query}. We also used an implementation of min-cost max-flow to compute the query-optimal offline arrangement of a two-choice linear-probing hash table of size 100,000 (at 100\% full, and with a fixed upper bound of $60$ probes per query), and found that the offline optimal number of probes per query is also $\approx 2.2$. These experiments are part of a larger experimental study that we plan to publish as a companion paper to this one.

\paragraph{Note on AI Use. }We remark that AI was not used in any way in preparing this paper, except for formatting and to catch/fix typos.

\section{Related Work}\label{sec:related-work}
\paragraph{Linear probing.} Linear probing was first introduced in 1954, when Amdahl, McGraw, and Samuel discovered it in the process of implementing symbol tables for a compiler on the IBM 701 machine \cite{Knuth98Vol3}. The hash table quickly became widely used within IBM, and was subsequently introduced to the public in a 1957 paper by Peterson \cite{Peterson57}. The first analysis of linear probing was in an unpublished note by Knuth \cite{knuth1963notes} in 1962, and the first published analysis was due to Konheim and Weiss \cite{KonheimWe66} in 1966. The observation that the average query time is unchanged by reordering elements within runs was due to Peterson \cite{Peterson57}, and was applied by Amble and Knuth \cite{AmbleKn74} to conclude that ordered linear probing hash tables exhibit $O(\epsilon^{-1})$ worst-case expected queries. Much more recently, several works \cite{bender2022linear,bender2024tight} have shown that the use of tombstones in an ordered-linear probing hash table naturally reduces the \emph{amortized} expected time of insertions to $\tilde{O}(\epsilon^{-1})$, and that such hash tables can further be modified to achieve amortized expected time $O(\epsilon^{-1})$ for all operations.

Most work on the subject (including this paper) assumes that the hash table has access to a uniformly random and fully independent hash function. However, several works have also shown how to analyze linear probing with more realistic hash functions, including $O(1)$-independent hash functions \cite{PaghPaRu07} and variations of tabulation hashing \cite{patracscu2012power,thorup2009string,thorup2012tabulation,puatracscu2013twisted,bercea2023locally}.

\paragraph{The power of two choices.} Our results are closely related (although only aesthetically) to the so-called power-of-two-choices phenomenon that has been studied in the load-balancing literature. This phenomenon considers a sequence of ball insertions, where each ball $x$ must go to one of two random bins $h_1(x)$ and $h_2(x)$. A classical result due to Azar et al.~\cite{azar1994balanced} says that, if each ball is placed in the emptier of its two choices, then the maximum bin load after $n$ insertions is $O(\log \log n)$, an exponential improvement over the classical bound of $O(\log n / \log \log n)$ that a single hash function would achieve.

This two-choice load-balancing result has subsequently been extended in many different ways. It was shown that an analogous result holds in the \emph{heavily loaded} setting, where $m \gg n$ balls are inserted, in which case the maximum load is $m/n + O(\log \log n)$ \cite{BerenbrinkCzSt00,BerenbrinkCzSt06,PeresTaWi10OnePlusBeta,TalwarWi13BalancedAllocations,los2022balanced}. The result has also been extended to settings where balls have weights \cite{talwar2014balanced, weighted2, PTW10, weighted4}, to dynamic settings where elements can be inserted and deleted \cite{Mitzenmacher00thepower,ColeMaMi98BallsDeletions,Vocking99,Vocking03,bender2021tiny,bansalkuszmaul22}, to batch-arrival settings \cite{parallel1, parallel2, parallel3, parallel4, parallel5}, to settings in which balls hash to edges on a graph of bins \cite{KP06,bansal2021well,bansalkuszmaul22}, and to settings where each ball has $d > 2$ choices for where to go \cite{Vocking03}, among others.

Our results can be viewed as a variation of the power-of-two choices phenomenon, but for linear-probing hash tables. From this perspective, our results come with several interesting surprises, including that, unlike the load-balancing setting, the optimal insertion algorithms appear to be \emph{non-greedy} approaches.

\paragraph{Two-choice hashing.} In the context of hash tables, the main area where power-of-two-choice ideas have appeared is in the study of cuckoo hashing \cite{pagh2008uniform} and its variations \cite{fotakis2005space,dietzfelbinger2007balanced}. Most closely related to our work is the study of \emph{bucketized cuckoo hashing} \cite{dietzfelbinger2007balanced}, where each element $x$ chooses between two buckets of some size $b$, and where elements can be moved around over time. A classical result, due to Dietzfelbinger and Weidling \cite{dietzfelbinger2007balanced}, says that buckets of size $b = \Theta(\log \epsilon^{-1})$ suffice to support load factors up to $1 - \epsilon$ (although achieving this bucket size with $\poly(\epsilon^{-1})$ insertions remains open \cite{dietzfelbinger2007balanced,frieze2018balanced,kuszmaul2025efficient}). This means that bucketized cuckoo hashing can be used to obtain a (worst-case!) query time of $O(\log \epsilon^{-1})$.

It is interesting to consider our results in comparison to those achieved by bucketized cuckoo hashing. Our results in Theorem \ref{thm:non-greedy} say that, if one is willing to use two-choice linear probing instead of cuckoo hashing (therefore settling for expected query time bounds, rather than worst-case query time bounds), then even \emph{without evictions} one can achieve a query time of $O(\log \epsilon^{-1})$. This comes in stark contrast with bucketized cuckoo hashing, where the insertion algorithm relies heavily on evictions. Theorem \ref{thm:non-greedyevictions}, on the other hand, says that if one \emph{is} willing to perform evictions, then the expected query time drops all the way to $O(1)$. Here, again, our result contrasts with bucketized cuckoo hashing where the optimal query time is $\Theta(\log \epsilon^{-1})$, even in the offline setting. It is also worth noting that all of our results come with efficient insertions, while the problem of supporting $\poly(\epsilon^{-1})$ insertions in bucketized cuckoo hashing (and with $b = O(\log \epsilon^{-1})$ buckets) remains open.

In addition to cuckoo hashing, two-choice linear probing (the subject of this paper) has also been studied, but with a different goal \cite{malalla2004two,dalal2023two}. Previous work has focused not on the expected time as a function of the load factor $1 - \epsilon$, but instead on the maximum query time (across the entire hash table) under the assumption that the load factor is $1 - \Theta(1)$. Here, it has been shown how to leverage classical two-choice load-balancing results in order to achieve a maximum query time of $O(\log\log n)$. The idea of using two-choice linear probing to improve the expected operation time, although briefly suggested in several experimental works \cite{nijalingappavertical,gurumurthy2018simd}, does not appear to have been explored in the theory literature until now.

\section{Preliminaries}\label{sec:prelims}

Throughout the paper, we will operate in the following setting: a sequence of elements are inserted into a hash table, where the table itself is an array of $n$ slots. Slots that do not contain elements are said to be \emph{free}, and for any free slot, that slot along with the maximal (possibly empty) contiguous sequence of non-free slots that immediately precedes it are said to form a \emph{run}. If the hash table currently has $(1 - \epsilon)n$ elements, it is said to have \emph{load factor} $1 - \epsilon$.

When discussing hashes of elements, we will typically use $h_i(u) < h_i(v)$ to mean that position $h_i(v)$ appears to the right of position $h_i(u)$ in the hash table. This convention is a slight abuse of notation, however, since the hash table uses wraparound (so position $0$ comes immediately after position $n - 1$). In practice, however, it will be clear from context how to interpret $h_i(u) < h_i(v)$ in the presence of wraparound.

Each inserted element $u$ chooses between two uniformly random hash functions $h_1(u)$ and $h_2(u)$ that it can use, and is then stored in the hash table using linear probing. We will often use $h(u)$ as a shorthand for whichever hash function $h_i(u)$ a given element is using. In all of the implementations that we consider (except for the warmup algorithm in Section \ref{sec:nongreedywarmup}), the hash table is \emph{ordered}, meaning that the elements $u$ are stored in increasing order of $h(u)$ (with wraparound handling as needed). This means that a \emph{query} for an element $u$ can simply follow the probe sequences $h_1(u), h_1(u) + 1, \ldots \pmod{n}$ and $h_2(u), h_2(u) + 1, \ldots \pmod{n}$ until either it finds the $u$, or until, in each probe sequence, it has found an element $v$ with $h(v) > h(u)$.\footnote{In several of our constructions, the query will also be able to stop examining a probe sequence if it encounters a free slot -- but, in others of our constructions, the query will not be able to stop examining a probe sequence until it encounters an element $v$ with $h(v) > h(u)$.}

For convenience, we will assume throughout the paper (except in the analysis of greedy in Section \ref{sec:greedy}) that the algorithms we design can keep track of $O(1)$ bits of metadata per element. This assumption is straightforward (but tedious) to remove in an essentially blackbox fashion, since such information can easily be encoded in the relative order of elements within blocks of size $O(1)$. As an example application of this assumption, we will assume throughout that we know for each element $x$ which of its two hashes $h_1(x), h_2(x)$ it is currently considered to be using (even though, in some very rare cases, those two hashes could be so close that we cannot tell which is being used simply by looking at $x$'s position).

Finally, it will often be helpful to talk about events that occur \emph{with high probability}. We follow the convention that an event occurs with high probability in $n$ if its probability is at least $1 - O(1/n^{c})$ for a large positive constant $c$ of our choice.

\section{A New Proof of a Classic Result}\label{sec:classic}

In this section, we give a new proof of a classic result, namely that (single-choice) linear probing, when filled to load factor $1 - \epsilon$, achieves insertion time $O(\epsilon^{-2})$. This proof will serve as the high-level starting point for our analysis of greedy two-choice linear probing in Section \ref{sec:greedy}. It is likely that the proof will also be of independent interest to the community. The proof is, as far as we know, the simplest proof to result in a sharp bound of $\epsilon^{-2}/2 + O(1)$ on the insertion time; and it is also the first proof to explicitly capture the way that the hash table evolves over time (as opposed to analyzing it as a history-independent data structure).

\begin{theorem}[\cite{Knuth63}]
Consider a classical linear-probing hash table on $n$ slots. For $\epsilon \le 1/2$, if we perform $(1 - \epsilon)n$ insertions, then the final insertion takes expected time $O(\epsilon^{-2})$.
\label{thm:classic}
\end{theorem}

As insertions are performed, we will keep track of two objects:
\begin{itemize}
    \item \defn{The Run Bag:} The run bag, denoted by $\mathcal{R}$, is the set of runs that are currently present in the hash table, where runs are as defined in Section \ref{sec:prelims}: a free slot together with the maximal (possibly empty) sequence of occupied slots immediately preceding it. Note that each run contains exactly one free slot, namely its final slot; in particular, a free slot that is immediately preceded by another free slot forms a run of length $1$ by itself. Thus runs are in one-to-one correspondence with free slots, and after $j$ insertions there are exactly $n - j$ runs. The run bag keeps track of the lengths of the runs, but not their positions.
    \item \defn{The Potential: } The potential $\Phi$ is defined  by
    $$\Phi = \sum_{R \in \mathcal{R}}  |R|^2/n.$$
\end{itemize}

A key observation in the analysis is that, if we condition on the current run bag $\mathcal{R}$, the \emph{relative order} in which the runs appear is \emph{uniformly random}. This is because the insertion process reveals no information about which runs are next to one another (we will formalize this more rigorously in Lemma \ref{lem:uniform-order} in Section \ref{sec:greedy}). Thus, for a given pair of runs $A, B \in \mathcal{R}$, the probability that $B$ comes directly after $A$ is simply $\frac{1}{|\mathcal{R}| - 1}$. This symmetry will be the key to our analysis.

The main step for proving Theorem \ref{thm:classic} is to establish the following lemma:
\begin{lemma}
Consider the $i$-th insertion. Let $\mathcal{R}$ (resp.~$\mathcal{R}'$) be the run bag before (resp.~after) the insertion, and let $\Phi$ (resp.~$\Phi'$) be the potential before (resp.~after) the insertion. Then,
$$\E[\Phi' \mid \Phi, \mathcal{R}] \le \Phi \cdot (1 + 2/(n - i)).$$
\label{lem:classicphi}
\end{lemma}
\begin{proof}
If the inserted element hashes into a run $A$, it probes forward to the free slot at the end of $A$ and fills it; $A$ then has no free slot of its own, and so merges with the run $B$ that follows it. Since the hash is uniformly random, and since the relative order of runs is uniformly random, one can think of $\mathcal{R}'$ being obtained from $\mathcal{R}$ as follows:
\begin{itemize}
    \item We sample a run $A \in \mathcal{R}$ by sampling each run with probability $|A|/n$.
    \item We sample a run $B \in \mathcal{R}\setminus\{A\}$ uniformly at random.
    \item We set $\mathcal{R}':= \mathcal{R} \setminus \{A, B\} \cup \{C\}$, where $C$ has length $|A| + |B|$.
\end{itemize}
This results in
$$\Phi' = \Phi - |A|^2/n - |B|^2/n + |C|^2/n = \Phi + 2|A| \cdot |B| / n.$$
Thus, it suffices to show that $\E[|A| \cdot |B| \mid \Phi, \mathcal{R}] \le \Phi \cdot \frac{n}{n - i}$.

By construction,
$$\E[|A| \mid \Phi, \mathcal{R}] = \sum_{R \in \mathcal{R}} |R| \cdot \Pr[A = R] = \sum_{R \in \mathcal{R}} |R| \cdot \frac{|R|}{n} = \Phi.$$
Moreover, conditioned on $A$, $B$ is uniformly random from $\mathcal{R} \setminus \{A\}$, so
$$\E[|B| \mid \Phi, \mathcal{R}, A] = \sum_{R \in \mathcal{R} \setminus \{A\}} \frac{|R|}{|\mathcal{R}| - 1} \le \frac{n}{|\mathcal{R}| - 1} = \frac{n}{n - i}.$$
Here, the second to final step uses the fact that the runs partition the $n$ slots; and the final step uses the fact that, prior to the $i$-th insertion, $i - 1$ elements are present, so there are $|\mathcal{R}| = n - i + 1$ runs.
It follows that $\E[|A| \cdot |B| \mid \Phi, \mathcal{R}] \le \Phi \cdot \frac{n}{n - i},$ as desired.
\end{proof}

From Lemma \ref{lem:classicphi}, we can conclude that the expected value of $\Phi$ increases by at most a factor of $(1 + 2/(n - i))$ during the $i$-th insertion. Applying this repeatedly, the overall expected value of $\Phi$ prior to the $(1 - \epsilon)n$-th insertion is at most
\begin{align*}
\E[\Phi] \,    &  \le \prod_{i = 1}^{(1 - \epsilon)n - 1} \left(1 + \frac{2}{n - i}\right).
\end{align*}
Using the identity $1 + \delta \le e^{\delta}$, this is at most
\begin{align*}
    & \exp\left( \sum_{i = 1}^{(1 - \epsilon)n - 1} 2/(n - i)\right) \\
    & = \exp\left( \sum_{i = \epsilon n + 1}^{n - 1} 2/i \right) \\
    & \le \exp\left(\int_{i = \epsilon n}^{n} \frac{2}{i} \, di\right) \\
    & = \exp\left(2 \ln n - 2 \ln (\epsilon n) \right) \\
    & \le \epsilon^{-2}.
\end{align*}
The fact that $\E[\Phi] \le \epsilon^{-2}$ implies that the expected number of probes made by the $(1 - \epsilon)n$-th insertion is at most $\epsilon^{-2}/2 + O(1)$, which implies Theorem \ref{thm:classic}.

\section{Greedy Two-choice Hashing}
\label{sec:greedy}

Define the \defn{greedy two-choice insertion policy} as follows: the insertion of a key $u$ considers the two probe sequences $h_1(u), h_1(u) + 1, h_1(u) + 2, \ldots \pmod n$ and $h_2(u), h_2(u) + 1, h_2(u) + 2, \ldots \pmod n$, and chooses between $h_1$ and $h_2$ based on which probe sequence has a free slot first (breaking ties towards, say, $h_1$). The insertion then places the element $u$ using whichever $h_i$ it chose, while maintaining the invariant that the elements $v$ within each run are stored in sorted order by their (chosen) hash. That is, if for each element $v$ we use $h(v)$ to denote the hash $h_j(v)$ that $v$ chose, then the elements $v$ in each run are in increasing order of $h(v)$.\footnote{In principle, there could exist a $v$ where we cannot easily tell which hash $v$ chose, because its two hashes are so close together. In this case, we follow the convention that $v$ is treated as using the larger of its two options $h_1(v)$ and $h_2(v)$.} Finally, queries to keys $u$ also examine the probe sequences $h_1(u), h_1(u) + 1, h_1(u) + 2, \ldots \pmod n$ and $h_2(u), h_2(u) + 1, h_2(u) + 2, \ldots \pmod n$, and stop when they either find $u$, or when they have determined that $u$ is not present (meaning that, in each probe sequence, the query has either seen a free slot or a key $v$ with $h(v) > h_i(u)$).\footnote{Here again, there could technically be a $v$ where we cannot tell which hash $v$ chose, and therefore what $h(v)$ is -- here, again, we follow the convention that $v$ is treated as using the larger of its two options $h_1(v)$ and $h_2(v)$.}

The main result of this section is the following theorem:
\thmGreedy*

Define the run bag $\mathcal{R}$, as in Section \ref{sec:classic}, to be the set of runs that are currently present in the hash table, where each run is represented only by a length, with no information about the relative order in which the runs appear. We begin by formalizing the observation that, conditioned on the current run bag $\mathcal{R}$, the relative order in which the runs appear is uniformly random. This proof also applies to the single-choice case.
\begin{lemma} \label{lem:uniform-order}
    Consider the hash table after $t$ insertions using the greedy two-choice policy, and let $\R$ be any run bag.
    Conditioned on the fact that the collection of runs in the hash table is equal to $\R$, all the cyclic permutations of the runs are equally likely for the hash table.
\end{lemma}
\begin{proof}
    We prove that the claim holds for any two-choice policy that is stable, i.e.\ the algorithm keeps using the same hash function for an element after it is inserted,
    and locally-determined, i.e.\ to choose between two hash values $r_1$ and $r_2$, the policy considers only the length of the run in which $r_i$ lies, and the position of the hash value in the run. Thus the proof applies equally well to the two-choice greedy algorithm or to the classical single-choice case.

    The claim can be proved by induction on $t$.
    The base case, $t = 0$, is trivial since the only valid run bag consists of $n$ runs of length $1$, which has only one cyclic permutation. We now prove the lemma for any $t > 0$, assuming by induction that it already holds for $t - 1$.

    Let $\hat{\R}$ be the run bag after $t$ insertions, and let $A$ and $B$ be the runs that are merged during the $t$-th insertion, with $A$ appearing before $B$ in the cyclic order of the runs. Also, let $\hat{\R}'$ be the run bag after the $t - 1$-th insertion, so $\hat{\R}'$ is the same as $\hat{\R}$ except that $\hat{\R}$ replaces $A$ and $B$ with a single run of length $|A| + |B|$. We will prove that, conditioned on $\mathcal{R}, A, B$, every cyclic permutation of the run bag $\hat{\R}$ is equally likely.

    Now consider a cyclic permutation $\pi$ of the run bag $\R$ after $t$ insertions. Note that, for each such $\pi$, there is a unique cyclic permutation $\pi'$ of the run bag $\hat{\R}'$ such that, if we condition on $\mathcal{R}, A, B$, the cyclic permutation $\pi$ occurs after the $t$-th insertion if and only if the cyclic permutation $\pi'$ occurs after the $(t - 1)$-th insertion. Therefore, $\Pr[\pi \mid \mathcal{R}, A, B] = \Pr[\pi' \mid \mathcal{R}, A, B] = \Pr[\pi' \mid \mathcal{R}', A, B]$. By induction, if we condition on $A, B$ being adjacent after insertion $t - 1$, then every valid permutation $\pi'$ is equally likely, occurring with probability $1 / (n - t)!$. It follows that the probability $\Pr[\pi \mid \mathcal{R}, A, B] = 1/(n - t)!$ is the same for all cyclic permutations $\pi$ of the run bag $\R$, which completes the proof.
\end{proof}

\begin{lemma}
    Define the \emph{configuration} of the hash table to be the bitmap indicating which slots are filled versus empty. Consider any configuration $H_0$ of the slots after $t$ insertions, and a sequence of $k$ insertions that follow, denoted by pairs of hash values $(r^{(1)}_1, r^{(1)}_2), \ldots, (r^{(k)}_1, r^{(k)}_2)$.
    Let $H_1$ be the configuration after the $k$ insertions, and let $H_2$ be the configuration after making all but the first of the $k$ insertions.
    Then, $H_1$ and $H_2$ differ only in one slot.
    \label{lem:lipshitz}
\end{lemma}
\begin{proof}
    We remark that the claim holds trivially for $k = 1$.
    Inserting an element into $H_0$ fills exactly one previously empty slot, and the rest of the configuration remains unchanged.
    Then, to prove the claim for $k > 1$, we argue that insertion is, in a sense, a Lipschitz transformation.
    That is, given two configurations $H$ and $H'$ that differ only in one slot, after making an insertion to each of them using the same hash values $(r_1, r_2)$,
    the resulting configurations also differ only in one slot.

    Next, we prove the Lipschitz property.
    Assume by symmetry that $H$ has more empty slots than $H'$.
    Let $\ell_1$ and $\ell_2$ denote the distance to the first empty slots following $r_1$ and $r_2$ in $H$.
    Let $A_1$ and $A_2$ denote the runs containing $r_1$ and $r_2$ in $H$, and let $B_1$ and $B_2$ be the following runs ($\ell'_1$, $\ell'_2$, $A'_1$, $A'_2$, $B'_1$, and $B'_2$ are defined similarly for $H'$).
    Assume by symmetry that $r_1$ is chosen in $H$ for the insertion, i.e.\ $\ell_1 \leq \ell_2$.
    Let $\hat{H}$ and $\hat{H}'$ denote the corresponding configurations after the insertion.
    We consider three cases, based on the location of the slot where $H$ and $H'$ differ, denoted by $p$.

    First, consider the case where $p$ is located at the end of $A_1$.
    That is, $H'$ is the same as $H$, except $A_1$ and $B_1$ are joined to create $A_1'$.
    We have assumed that in $H$, $r_1$ is chosen for insertion, i.e.\ $p$ is filled.
    Therefore, $\hat{H}$ is equal to $H'$.
    Hence, $\hat{H}'$ differs from $\hat{H}$ in only one slot, since it is obtained by filling an empty slot of $H'$.

    Second, consider the case where $p$ is located at the end of $A_2$.
    That is, $H'$ is the same as $H$, except $A_2$ and $B_2$ are joined to create $A_2'$.
    In this case, it holds that $\ell'_2 \geq  \ell_2 \geq \ell_1 = \ell_1'$.
    As a result, $r_1$ is selected for insertion also in $H'$.
    Therefore, $H$ and $H'$ fill the same empty slot, and $\hat{H}$ and $\hat{H}'$ differ only in one slot, namely $p$.

    Finally, consider the case where $p$ is located neither at the end of $A_1$ nor at the end of $A_2$.
    In this case, it holds that $\ell_1 = \ell_1'$ and $\ell_2 = \ell_2'$.
    Therefore, similar to the previous case, $r_1$ is chosen for insertion also in $H'$, and $\hat{H}$ and $\hat{H}'$ differ only in one slot.
    This concludes the proof of the Lipschitz property, and in turn the proof of the claim.
\end{proof}

\begin{corollary}
    Consider any configuration $H_0$ of the slots after $t$ insertions, and a sequence of $k$ insertions that follow, denoted by pairs of hash values $(r^{(1)}_1, r^{(1)}_2), \ldots, (r^{(k)}_1, r^{(k)}_2)$.
    Let $H_1$ be the configuration after the $k$ insertions, and let $H_2$ be the configuration after the same sequence of $k$ insertions but with some pair $(r^{(i)}_1, r^{(i)}_2)$ replaced with an alternative $({r^{(i)}_1}', {r^{(i)}_2}')$.
    Then, $H_1$ and $H_2$ differ in at most 2 slots.
    \label{cor:lipshitz}
\end{corollary}

Throughout the section, we will consider an arbitrary \emph{truncation threshold} $T \in n^{\Omega(1)} \cap O(n^{1/5})$. We define for a run $R$ the \emph{truncated length} of $R$ to be $\overline{\card{R}} = \min(\card{R}, T)$, and we define both the \emph{truncated potential} and the \emph{long potential} functions for a run bag $\R$ by:
    $$
    \Phi_T = \sum_{R \in \R} \card{R} \cdot \overline{\card{R}} / n
    \qquad \textnormal{and} \qquad
    \Phi_L = \sum_{\substack{R \in \R \\ \card{R} > T}}
    \card{R}^2 /n.
    $$
    We will also let $\mathcal{L} := \{R \in \R \mid \card{R} > T \}$ be the set of \emph{long runs}, and $\vol(\mathcal{L}) := \sum_{R \in \mathcal{L}} \card{R}$ be their total \emph{volume}. Note that all of these variables ($\mathcal{R}, \mathcal{L}, \Phi_T, \Phi_L$) implicitly depend on the point in time that we are analyzing. Also, note that, since $T \ge n^{\Omega(1)}$, and since $\epsilon^{-1} \le n^{o(1)}$ (by assumption), we have that $\poly(\epsilon^{-1}) = T^{o(1)}$, a fact that we will use to simplify our algebra throughout.

    Intuitively, $\Phi_L$ should be $0$ (we expect no runs of length $\ge T$), while $\Phi_T$ should be exactly equal to $\Phi$ as it was defined in the previous section. However, this intuition turns out to be remarkably difficult to formalize, and instead we take the following approach in the analysis: First, we analyze $\Phi_T$ on its own, making aggressive use of the fact that each individual run contributes at most $O(n^{2/5})$ to it (namely, this allows for the concentration bound in Lemma \ref{lem:McDiarmid}). Then, having bound $\Phi_T$, we show how to obtain a weak but ``good-enough'' bound for $\Phi_L$ by directly relating $\Phi_L$ both to itself and to $\Phi_T$.

    We begin by establishing a concentration bound on $\Phi_T$. This bound will be important later in our analysis, because it means that for any event $E$ which occurs with probability $\Omega(1)$, we have $\E[\Phi_T \cdot \mathbf{1}[E]] = \Omega(\E[\Phi(T)])$, a fact that is not necessarily true a priori.
\begin{lemma}
    Consider any configuration $H_0$ of the slots after $t$ insertions, and a sequence of $k$ insertions that follow, denoted by pairs of hash values $(r^{(1)}_1, r^{(1)}_2), \ldots, (r^{(k)}_1, r^{(k)}_2)$. Then, if we consider $\Phi_T$ after the insertions are completed, we have $\Phi_T \in (1 \pm o(1))\E[\Phi_T]$ with high probability in $n$. Moreover, if we let $a \le b \in \mathbf{N}$, and let $I_{a, b}$ be the number of runs with lengths in $[a, b]$ after the insertions are completed, we have that $|I_{a, b} - \E[I_{a, b}]| \le \tilde{O}(\sqrt{n})$ with high probability in $n$.
    \label{lem:McDiarmid}
\end{lemma}
\begin{proof}
    We prove this lemma with the help of McDiarmid's inequality \cite{McDiarmid89}, a variation on Azuma's inequality which states that: for any function $F(X_1, X_2, \ldots, X_m)$ of independent random variables, if changing a given $X_i$ never changes the function by more than $C$, then $\Pr[|F - \E[F]| \ge k \sqrt{m} C] \le e^{-\Omega(k)}$.

    Having fixed $H_0$, we can express $\Phi_T$ as a function $f((r^{(1)}_1, r^{(1)}_2), \ldots, (r^{(k)}_1, r^{(k)}_2))$ of $k \le n$ independent random variables. By Corollary \ref{cor:lipshitz}, changing any one of the inputs to $f$ changes the output by at most $O(T^2/n) = O(n^{-3/5})$. It follows by McDiarmid's inequality that $\Pr[|\Phi_T - \E[\Phi_T]| \ge k \sqrt{n} \cdot n^{-3/5}] \le e^{-\Omega(k^2)}$, which implies that $|\Phi_T - \E[\Phi_T]| \le 1/n^{\Omega(1)}$ with high probability in $n$. Since $\Phi_T \ge 1$ (deterministically), it follows that $\Phi_T \in (1 \pm o(1))\E[\Phi_T]$ with high probability in $n$.

    Likewise, having fixed $H_0$, we can express $I_{a, b}$ as a function $g((r^{(1)}_1, r^{(1)}_2), \ldots, (r^{(k)}_1, r^{(k)}_2))$ of $k \le n$ independent random variables. By Corollary \ref{cor:lipshitz}, changing any one of the inputs to $g$ changes the output by at most $O(1)$. It follows by McDiarmid's inequality that $\Pr[|I_{a, b} - \E[I_{a, b}]| \ge k \sqrt{n}] \le e^{-\Omega(k^2)}$, from which the corollary follows.
\end{proof}

Having established a concentration bound on $\Phi_T$, we now turn to bounding its growth on a given insertion.

\begin{lemma}
    If a run $A$ is selected for insertion, which is followed by a run $B$, then as a result of the insertion $\Phi_T$ grows by at most $2 \overline{\card{A}} \card{B} / n$.
    \label{lem:phi-t-growth}
\end{lemma}
\begin{proof}
    Let $\Phi_T'$ be the truncated potential after the insertion.
   We consider four cases based on $|A|$ and $|B|$, and examine in each case the contribution of $A$ and $B$ to $\Phi_T$ versus the contribution of the joined run to $\Phi'_T$.

    \begin{enumerate}
        \item{$\card{A} \leq T$ and $\card{B} \leq T$:}
        $$
        \Phi_T' - \Phi_T \leq \left((\card{A} + \card{B})^2 - \card{A}^2 - \card{B}^2\right) / n
        = 2\card{A}\card{B}/n.
        $$

        \item{$\card{A} \leq T$ and $\card{B} > T$:}
        $$
        \Phi_T' - \Phi_T = \left((\card{A} + \card{B}) \cdot T - \card{A}^2 - \card{B} \cdot T\right) / n
        = \card{A} \cdot (T - \card{A}) / n.
        $$

        \item{$\card{A} > T$ and $\card{B} \leq T$:}
        $$
        \Phi_T' - \Phi_T = \left((\card{A} + \card{B}) \cdot T - \card{A} \cdot T- \card{B}^2\right) / n
        = (T - \card{B}) \cdot \card{B} /n
        $$

        \item{$\card{A} > T$ and $\card{B} > T$:}
        $$
        \Phi_T' - \Phi_T \leq \left((\card{A} + \card{B}) \cdot T - \card{A} \cdot T - \card{B} \cdot T\right) / n
        = 0.
        $$
    \end{enumerate}

    It can be seen that all four values are smaller than $2\overline{\card{A}}\card{B}/n$.
\end{proof}

When an insertion is performed, the run $A$ in Lemma \ref{lem:phi-t-growth} is chosen from two runs $A_1$ and $A_2$ (containing the hash values $h_1(u)$ and $h_2(u)$, respectively). Roughly speaking, to prove that greedy-two choice insertions beat greedy single-choice insertions, we need to show that $\E[\min(|A_1|, |A_2|)] \le (1 - \Omega(1)) \E[|A_1|]$.

If we want to show that $\min(|A_1|, |A_2|)$ tends to be significantly smaller than $|A_1|$, then we need to show that there is a sense in which the run bag $\mathcal{R}$ is imbalanced. Let $\gamma \in (0, 1/4)$ be a small positive constant (whose value will be determined later). Say that a run bag $\R$ is \emph{imbalanced} if there is a length $\ell$ such that at least a $\gamma$ fraction of the table is occupied by runs of truncated length at most $\ell$, and at least a $\gamma$ fraction is occupied by runs of truncated length at least $2 \ell$. Otherwise, the run bag is \emph{balanced}. Note that, if a bag is balanced, then it has the following natural property:

\begin{lemma}
    If a run bag is balanced, then there exists a length $\ell$ such that at least a $1/4$-fraction of the table is occupied by runs of truncated length between $[\ell, 2 \ell]$.
    \label{lem:balanced-lemma1}
\end{lemma}
\begin{proof}
We will prove the contrapositive: If for all $\ell$, less than a $1/4$ fraction of the table is occupied by runs of truncated length in $[\ell, 2\ell]$, then the run bag is not balanced.

For $i \ge 0$, let $\vol^-_i$  be the total volume occupied by runs of truncated length at most $2^i$,
    and let $\vol^+_i$ be the total volume occupied by runs of truncated length at least $n / 2^i$.
    Let $i^*$ be the smallest index such that $\vol^-_{i^*} \ge n/2$. Then, since no interval contributes more than $n/4$ to the volume, we have that $\vol^-_{i^* - 1} \ge n/4$, that $\vol^-_{i^*} \le 3n/4$, and therefore that $\vol^+_{i^*} \ge n/4$.  It follows that at least a $1/4$ fraction of the table is in runs of length at most $2^{i^* - 1}$ and at least a $1/4$ fraction of the table is in runs of length at least $2^{i^*}$, so the run bag is imbalanced.
\end{proof}

It turns out that we will not be able to prove that the run bag is \emph{always} imbalanced. (It is not, for example, imbalanced initially.) Instead, we can show that if there is some moment in time when it is balanced, then there is some moment in the near future when it will be imbalanced (or when $\vol(\mathcal{L})$ will be unexpectedly large). Note that, in the following lemma we do not make any effort to optimize the constants involved, and instead optimize for readability only.
\begin{lemma}
Suppose that, after the $(1 - \epsilon)n$-th insertion, the probability of the run bag being balanced is at least $\Omega(1)$. Then, after the $(1 - 0.9\epsilon)n$-th insertion, one of the following two statements are true:
\begin{enumerate}
    \item With probability $\Omega(1)$, the run bag is imbalanced;
    \item With probability $\Omega(1)$, $\vol(\mathcal{L}) \ge \Omega(n)$.
\end{enumerate}
\label{lem:balanced-lemma2}
\end{lemma}
\begin{proof}
Condition on the run bag $\mathcal{R}$ after the $(1 - \epsilon)n$-th insertion. Suppose that $\vol(\mathcal{L}) \le n/8$, and that $\mathcal{R}$ is balanced. By Lemma \ref{lem:balanced-lemma1}, there exists some $\ell \le T$ such that at least a $1/4$ fraction of the table is occupied by runs of truncated lengths in $[\ell, 2\ell]$. Since $\vol(\mathcal{L}) \le n/8$, it follows that at least a $1/8$ fraction of the table is occupied by runs of (non-truncated) length between $[\ell, 2 \ell]$. Let $\mathcal{Q}$ be the set of such runs, and note that by construction $|\mathcal{Q}| \ge  n / (16 \ell)$.

Say that a run $B$ is \emph{merged from the left} during an insertion if the insertion combines the run $A$ before $B$ with $B$ to get a new run $C = A \circ B$. On each insertion, the run that gets merged from the left is uniformly random from $\mathcal{R} \setminus \{A\}$. Therefore, during each of the insertions $(1 - \epsilon)n + 1, \ldots, (1 - 0.9\epsilon)n$, each run has probability at most $\frac{1}{0.9 \epsilon n}$ of being merged from the left. The probability of a given run $R \in \mathcal{R}$ ever being merged from the left (or being part of a larger run that is merged from the left) during those insertions is therefore at most $(0.1 \epsilon n) / (0.9 \epsilon n) < 1/2$.

The expected number of runs in $\mathcal{Q}$ that \emph{do not} get merged from the left during the insertions $(1 - \epsilon)n + 1, \ldots, (1 - 0.9\epsilon)n$ is therefore at least $|\mathcal{Q}| / 2$. These runs are guaranteed to be separate from one another after insertion $(1 - 0.9\epsilon)n$ (i.e., they have not been merged with each other). Therefore, after step $(1 - 0.9\epsilon)n$, the expected number of runs with length $\ge \ell$ is at least $|\mathcal{Q}| / 2 \ge n / (32\ell)$. By Lemma \ref{lem:McDiarmid}, it follows that, with high probability in $n$, the number of such runs is at least $n / (64 \ell)$. Since there can be at most $n/(128\ell)$ runs of length $\ge 128 \ell$, this forces there to be at least
\begin{equation} n / (128 \ell)
 \label{eq:32ell}
\end{equation}
runs with lengths in $[\ell, 128 \ell]$.

So far we have shown that there are many runs with lengths in $[\ell, 128 \ell]$. To complete the proof, we will show that there are also many runs with lengths at least $256\ell$. Consider a run $R \in \mathcal{Q}$. We will next show that, with probability $\Omega(1)$, $R$ is in a run of length at least $256\ell$ after the $(1 - 0.9\epsilon)n$-th insertion.

During any insertion after $(1 - \epsilon)n$, we claim that the insertion uses run $R$ (i.e., the run containing $R$ serves the role of $A$ during the insertion) with probability at least $\Omega(\ell/n)$. In particular, for the element $u$ being inserted, there is at least an $\ell/(2n)$ probability that $u$'s first hash is in the final $\ell/2$ slots of $R$, and at least a $\vol(\mathcal{Q})/(2n)$ probability that the other hash is in a run of length $\ge \ell$ and is \emph{not} in that run's final $\ell/2$ slots. Recalling that $\vol(\mathcal{Q}) \ge n/8$ (by assumption), this gives an overall probability of at least
$$(\ell / (2n)) \cdot (n/8)/(2n) = \Omega(\ell/n)$$
of the run $R$ being used by the insertion.

If the run is used by the insertion, then the probability of it merging with another run $R'$ of length at least $\ell$ (remember, by \eqref{eq:32ell}, there are at least $\Omega(n / \ell)$ such runs!) is at least $\frac{\Omega(n/\ell)}{\epsilon n} = \Omega((\epsilon \ell)^{-1})$ (here, as a slight abuse, we ignore the $1/\poly(n)$ probability of \eqref{eq:32ell} failing, as this has negligible effect on the lemma). So, on each insertion, the probability that $|R|$ grows by at least $\ell$ is at least
$$\Omega(\ell/n) \cdot \Omega((\epsilon \ell)^{-1}) = \Omega((\epsilon n)^{-1}).$$
Since this is true independently for each insertion (again, ignoring the $1/\poly(n)$ probability of \eqref{eq:32ell} failing), the number of times that $|R|$ grows by at least $\ell$ is lower bounded (stochastically) by a binomial random variable $\operatorname{Bin}(0.1 \epsilon n, \Omega((\epsilon n)^{-1}))$. This binomial random variable (which has mean $\Omega(1)$) is at least 256 with probability $\Omega(1)$. Therefore, with probability $\Omega(1)$, $R$ is in a run of length at least $256 \ell$ after the $(1 - 0.9\epsilon)n$-th insertion. By Lemma \ref{lem:McDiarmid}, it follows that, with high probability in $n$, the volume of runs of length at least $256 \ell$ is at least $\Omega(n)$ after the $(1 - 0.9\epsilon)n$-th insertion.

At this point, we have that, with high probability in $n$: the volume of runs of length $\le 128 \ell$ is $\Omega(n)$, while the volume of runs of length $\ge 256 \ell$ is also $\Omega(n)$. Assuming the constant $\gamma$ in the definition of imbalanced is chosen appropriately, it follows that either the run bag is imbalanced, or that an $\Omega(1)$ fraction of the volume is in runs of length $\ge T$. Recalling our assumptions at the start of the proof, our analysis holds for any initial run bag $\mathcal{R}$ so long as the total volume of long runs was at most $n/4$, and so long as $\mathcal{R}$ was balanced. Of course, if the initial volume of the long runs was $\Omega(n)$, then the target of the lemma would hold trivially. So, the target goals of the lemma are true with high probability so long as the initial run bag was balanced, which itself occurs with probability $\Omega(1)$ by assumption. This completes the proof.
\end{proof}

We are now prepared to begin calculating $\E[\Phi_T]$ for the greedy two-choice insertion algorithm. We begin by formalizing an observation that will be helpful throughout:
\begin{lemma}
Consider the $i$-th insertion, let $\mathcal{R}$ be the run bag prior to the insertion, and let $A$ and $B$ be the two runs that the insertion merges. Then, no matter the value of $A$, the conditional expectation of $|B|$ is at most $\frac{n}{n - i}$.
\label{lem:boundB}
\end{lemma}
\begin{proof}
Once $A$ is selected, $B$ is uniformly random from $\mathcal{R} \setminus \{A\}$. Since $\mathcal{R} \setminus \{A\}$ has size $n - i$, and since $\sum_{R \in \mathcal{R} \setminus \{A\}} |R| \le n$, the lemma follows.
\end{proof}

We also formalize the way in which choosing between two runs changes the expected length of the run that the insertion uses:
\begin{lemma}
    Consider the insertion of an element $u$, let $A_1$ be the run that $h_1(u)$ is in, let $A_2$ be the run that $h_2(u)$ is in, and let $A \in \{A_1, A_2\}$ be the run that the insertion selects to use. Then, supposing that $|A_1| \le |A_2|$, we have that
    $$\E[|A|] = \frac{|A_1| + |A_2|}{2} - \frac{(|A_2| - |A_1|)^2}{2|A_2|}.$$
    \label{lem:minA}
    \end{lemma}
    \begin{proof}
    If $h_2(u)$ is in the final $|A_1|$ positions of $|A_2|$ (which happens with probability $|A_1| / |A_2|$), then $A$ is equally likely to be $A_1$ or $A_2$. But if $h_2(u)$ is not in the final $|A_1|$ positions of $|A_2|$ (which happens with probability $(|A_2| - |A_1|) / |A_2|$), then $A$ is guaranteed to be $A_1$. It follows that
    $$\E[|A|] = \frac{|A_1|}{|A_2|} \cdot \frac{|A_1| + |A_2|}{2} + \frac{|A_2| - |A_1|}{|A_2|} \cdot |A_1|.$$
    This, in turn, can be rewritten as
    $$\E[|A|] = \frac{|A_1| + |A_2|}{2} + \frac{|A_2| - |A_1|}{|A_2|} \cdot \left(|A_1| - \frac{|A_1| + |A_2|}{2}\right) = \frac{|A_1| + |A_2|}{2} - \frac{(|A_2| - |A_1|)^2}{2|A_2|}.$$
    \end{proof}

Next we prove that, regardless of whether $\mathcal{R}$ is imbalanced, we have a bound on $\Phi_T$'s growth rate that is the same as the bound proven in Section \ref{sec:classic} for the single-choice strategy.

\begin{lemma}
    Consider the $i$-th insertion. Let $\Phi_T$ be the truncated potential prior to the insertion, and let $\Phi_T'$ be the truncated potential after the insertion. Then,
    $$\E[\Phi_T'] \le \E[\Phi_T] \cdot (1 + 2/(n - i)).$$
    \label{lem:phi-t-growth-basic}
    \end{lemma}
    \begin{proof}
    Condition on the run bag $\R$ prior to the insertion, and on the value of $\Phi_T$. Let $|A_1|$ be the length of the run that $h_1(u)$ is in, and let $|A_2|$ be the length of the run that $h_2(u)$ is in. Finally, let $A \in \{A_1, A_2\}$ be the run that the insertion chooses to use. Note that by Lemma \ref{lem:minA}, $\E[\overline{|A|}] \le \E[(|A_1| + |A_2|)/2] = \E[|A_1|]$.

    By Lemma \ref{lem:phi-t-growth}, the expected increase to $\Phi_T$ is at most
    \begin{align*}
    \E[2 \overline{|A|} \cdot |B| / n].
    \end{align*}
    By Lemma \ref{lem:boundB}, this is at most
    \begin{align*}
            &= \frac{2}{n}\cdot\E[\overline{|A|}] \cdot \frac{n}{n - i} \\
            &\le \frac{2}{n} \cdot \E[\overline{|A_1|}] \cdot \frac{n}{n - i} \\
            &= 2 \Phi_T \cdot \frac{n}{n - i}. \\
    \end{align*}
    Since this holds conditioned on any fixed run bag $\R$ and value of $\Phi_T$, it follows that, even when we remove the conditioning, we have
    \begin{align*}
    \E[\Phi_T'] \le \E[\Phi_T] \cdot (1 + 2/(n - i)).
    \end{align*}
    \end{proof}

Now, critically, we consider the case where the run bag is imbalanced, and show that in this case we get a notably better bound on $\Phi_T$'s growth rate.
\begin{lemma}
Suppose that, prior to the $i$-th insertion, the run bag has probability at least $\Omega(1)$ of being imbalanced. Let $\Phi_T$ be the truncated potential prior to the insertion, and let $\Phi_T'$ be the truncated potential after the insertion. Then,
$$\E[\Phi_T'] \le \E[\Phi_T] \cdot \left(1 + (2 - \Omega(1))\frac{n}{n - i}\right).$$
\label{lem:phi-t-growth-imbalanced}
\end{lemma}
\begin{proof}
Condition on the run bag $\R$ prior to the insertion, and on the value of $\Phi_T$. Suppose that the run bag is imbalanced: there exists a length $\ell$ such that at least an $\Omega(1)$-fraction of the table volume is occupied by runs of truncated length at most $\ell$, and at least an $\Omega(1)$-fraction is occupied by runs of truncated length at least $2\ell$.

Let $u$ be the element being inserted. Let $A_1$ be the run that $h_1(u)$ is in, and let $A_2$ be the run that $h_2(u)$ is in. Let $A \in \{A_1, A_2\}$ be the run that the insertion selects to use.

The key to proving the lemma is to show that
\begin{equation}
\E[|A|] \le (1 - \Omega(1)) \E[|A_1|].
\label{eq:minA}
\end{equation}
By Lemma \ref{lem:minA}, this is equivalent to showing that
$$\E[(|A_1| - |A_2|)^2 / \min(|A_1|, |A_2|)] \ge \Omega(\E[|A_1|]),$$
for which it, in turn, suffices to show that
$$\E[\big||A_1| - |A_2|\big|] \ge \Omega(\E[|A_1|]).$$
By assumption, $\Pr[|A_1| \ge 2\ell] \ge \Omega(1)$, so $\E[\overline{|A_1|} \cdot \mathbf{1}[|A_1| \ge 2\ell]] \ge \Omega(\E[|A_1|])$. But, if $|A_1| \ge 2\ell$, then $|A_2| \le \ell$ with probability $\Omega(1)$, in which case $|A_1| - |A_2| \ge |A_1| / 2$. It follows that
$$\E[\big||A_1| - |A_2|\big|] \ge \Omega(\E[|A_1| \cdot \mathbf{1}[|A_1| \ge 2\ell]]) = \Omega(\E[|A_1|]),$$
which implies \eqref{eq:minA}.

Now, suppose that the insertion merges $A$ with run $B$. By Lemma \ref{lem:phi-t-growth}, the increase to $\Phi_T$ is at most $2 \overline{|A|} |B| / n$. The expected increase is therefore at most
\begin{align*}
\E[2 \overline{|A|} \cdot |B| / n]
    &\le \frac{2}{n} \E[|A|] \cdot \frac{n}{n - i} \tag{by Lemma \ref{lem:boundB}}\\
    &\le \frac{2}{n} (1 - \Omega(1)) \E[|A_1|] \cdot \frac{n}{n - i} \tag{by \eqref{eq:minA}}\\
    &= 2 \cdot  (1 - \Omega(1)) \Phi_T \cdot \frac{n}{n - i}.\\
    & = (2 - \Omega(1)) \Phi_T \cdot \frac{n}{n - i}.
\end{align*}

Throughout the above argument, we have conditioned on a specific imbalanced run bag $\R$ and therefore also on a specific value of $\Phi_T$. We know, however, that the value of $\Phi_T$ is by Lemma \ref{lem:McDiarmid} of the form $(1 \pm o(1)) \E[\Phi_T]$ with high probability. We also know that the probability of the run bag being imbalanced prior to insertion $i$ is at least $\Omega(1)$. Together, these facts imply that
$$\E[\Phi_T'] \le \E[\Phi_T] \cdot \left(1 + (2 - \Omega(1))\frac{n}{n - i}\right),$$
as desired.
\end{proof}

Before we can bound $\Phi_T$, we will need one more observation, which is that $\vol(\mathcal{L})$ is small with high probability -- this, in particular, will allow us to use Lemma \ref{lem:balanced-lemma2} without worrying about the case of $\vol(\mathcal{L}) = \Omega(n)$, and will also be important in our analysis later on of $\E[\Phi_L]$.
\begin{lemma} \label{lem:long-runs-small-volume}
    The total volume of the long runs is at most $O(n\epsilon^{-2} /T) = n / T^{1 - o(1)}$.
\end{lemma}
\begin{proof}
    We begin by proving that $\E[\Phi_T] = O(\epsilon^{-2})$, and then using that to deduce a bound on $\vol(\mathcal{L})$. By Lemma \ref{lem:phi-t-growth-basic}, the expected value of $\Phi_T$ after insertion $(1 - \epsilon)n$ is at most
    $$\prod_{j = 1}^{(1 - \epsilon)n - 1} (1 + 2/(n - j)),$$
    which by the same calculation as in Section \ref{sec:classic} is at most $O(\epsilon^{-2})$. By Lemma \ref{lem:McDiarmid}, it follows that $\Phi_T = O(\epsilon^{-2})$ with high probability in $n$.

    Now consider the contribution of $\mathcal{L}$ to $\Phi_T$. We have
    \begin{align*}
        \Phi_T &= \sum_{R \in \R} \card{R} \cdot  \overline{\card{R}} / n \\
    &\geq \sum_{R \in \mathcal{L}} \card{R} \cdot  \overline{\card{R}} / n \\
    &= \sum_{R \in \mathcal{L}} \card{R} \cdot T / n \\
    &= \vol(\mathcal{L}) \cdot T /n.
    \end{align*}
    Since $\Phi_T = O(\epsilon^{-2})$ with high probability in $n$, it follows that $\vol(\mathcal{L}) = O(n \epsilon^{-2} / T)$ with high probability in $n$.
\end{proof}

Putting the pieces together, we can now bound $\E[\Phi_T]$.
\begin{lemma}
Immediately before insertion $(1 - \epsilon)n$, the expected value of $\Phi_T$ is at most
$$\epsilon^{-(2 - \Omega(1))}.$$
\label{lem:phiT}
\end{lemma}
\begin{proof}
Now consider the $(1 - 1/a)n$-th and $(1 - 0.9/a)n$-th insertions, where $0.9/a \ge \epsilon$. By Lemma \ref{lem:balanced-lemma2}, at least one of these two insertions has the property that, prior to the insertion occurring the run bag is imbalanced with probability $\Omega(1)$. (Note that Lemma \ref{lem:balanced-lemma2} also allows for the possibility that $\vol(\mathcal{L}) = \Omega(n)$ with probability $\Omega(1)$, but we have already in Lemma \ref{lem:long-runs-small-volume} established this is not the case!)

Define $c_i$ to be $1$ if, prior to insertion $i$, the run bag has probability at least $\Omega(1)$ of being imbalanced, and $0$ otherwise. Then, by Lemmas \ref{lem:phi-t-growth-imbalanced} and \ref{lem:phi-t-growth-basic}, the expected value of $\Phi_T$ after insertion $i$ is at most
\begin{align*}
\E[\Phi_T] & \le \prod_{i = 1}^{(1 - \epsilon)n - 1} (1 + (1 - \Omega(c_i)) \cdot 2/(n - i)) \\
& \le \exp\left(\sum_{i = 1}^{(1 - \epsilon)n - 1} 2 / (n - i) - \sum_{i = 1}^{(1 - \epsilon)n - 1} \Omega(c_i / (n - i))\right).
\end{align*}
We have already shown that, for any $2a \le \epsilon^{-1}$, we have $c_{(1 - 1/a)n} + c_{(1 - 0.9/a)n} \ge 1$. This implies that $$\sum_{i = 1}^{(1 - \epsilon)n - 1} \Omega(c_i / (n - i)) \ge \Omega\left(\sum_{i = 1}^{(1 - \epsilon)n - 1} 1 / (n - i)\right).$$ Therefore,
$$\E[\Phi_T] \le \exp\left((1 - \Omega(1)) \cdot \sum_{i = 1}^{(1 - \epsilon)n - 1} 2 / (n - i)\right),$$
which by the same calculation as in Section \ref{sec:classic} is at most $O(\epsilon^{-(2 - \Omega(1))})$.
\end{proof}

\paragraph{Bounding $\Phi_L$.} Now we turn to bounding the expected value of $\Phi_L$.

Let $A^{(i)}$ and $B^{(i)}$ be the runs that are merged together by the $i$-th insertion. Define
$$\Psi(i) = (|A^{(i)}| + |B^{(i)}|)^2 \cdot \mathbf{1}[|A^{(i)}|, |B^{(i)}| < T \text{ and } |A^{(i)}| + |B^{(i)}| \ge T].$$
The next lemma shows that, if we ignore $\Psi(i)$, there is a nice way to bound the expected amount by which $\Phi_L$ grows on each insertion. Specifically, we are able to bound the growth of $\Phi_L$ as a function of $\Phi_L$, $\Phi_T$, and $T$. We remark that the argument that follows relies heavily on Lemma \ref{lem:long-runs-small-volume}, which tells us that $\vol(\mathcal{L}) = O(n / T^{1 - o(1)})$.

\begin{lemma}
Consider the $i$-th insertion, and let $u$ be the key being inserted.
Let $\Phi_T$ and $\Phi_L$ be the truncated and long potentials before the insertion,  and let $\Phi_T'$ and $\Phi_L'$ be the truncated and long potentials after the insertion. Then,
    $$(\E[\Phi_L'] - \E[\Phi_L]) \cdot n \le O(\E[\Phi_L] / T^{1/8}) + O(1/T^{1/4}) + O(\E[\Phi_T] / T^{1/2}) + \E[\Psi(i)].$$
    \label{lem:foob1}
\end{lemma}
\begin{proof}
Let $u$ be the key being inserted. Let $A_1$ be the run that $h_1(u)$ is in, let $A_2$ be the run that $h_2(u)$ is in, and let $A \in \{A_1, A_2\}$ be the run that is selected for insertion. Let $B$ be the run that appears directly after $A$, meaning that the insertion combines $A$ and $B$ into a new run.

Note that
\begin{align*}
(\Phi_L' - \Phi_L) \cdot n & = O(|A| \cdot |B| \cdot \mathbf{1}[|A| \ge T \text{ or } |B| \ge T]) + O((|A| + |B|)^2 \cdot \mathbf{1}[|A|, |B| < T \text{ and } |A| + |B| > T]) \\
& = O(|A| \cdot |B| \cdot \mathbf{1}[|A| \ge T \text{ or } |B| \ge T]) + \Psi(i).
\end{align*}
To bound this, we begin by bounding
$$\E[|A| \cdot |B| \cdot \mathbf{1}[|A| \ge T]).$$
By Lemma \ref{lem:boundB}, this is at most
\begin{equation}
\E[|A|  \cdot \mathbf{1}[|A| \ge T]] \cdot \frac{n}{n - i}.
\label{eq:ab-bound}
\end{equation}
Up to a factor of two, we may assume without loss of generality $A = A_1$. In order for $A = A_1$ and $|A| \ge T$ to occur, one of two things must happen:
\begin{enumerate}
    \item $|A_1| \ge T$, $|A_2| \ge \sqrt{T}$ and $A = A_1$: Let $T' = \sqrt{T}$, and let $\mathcal{L}'$ be the set of runs of length at least $T'$. Applying Corollary \ref{lem:long-runs-small-volume}, but for $T' = \sqrt{T}$, we know that we have $\vol(\mathcal{L}') = O(n / T'^{1 - o(1)}) = O(n / T^{1/4 - o(1)})$ with high probability in $n$. Therefore,
    $$\Pr[A_2 \in \mathcal{L}'] \le O(1/T^{1/4 - o(1)}).$$
    We can therefore bound the contribution of this case to \eqref{eq:ab-bound} by
    \begin{align*}& O(1/T^{1/4 - o(1)}) \cdot \E\left[\sum_{R \in \mathcal{L}} |R| \cdot \Pr[A_1 = R] \cdot \frac{n}{n - i} \right] \\
    &\le O(1/T^{1/4 - o(1)}) \cdot \E[\Phi_L] \cdot \frac{n}{n - i}  \\
    &\le O(\E[\Phi_L] / T^{1/8}),
   \end{align*}
    where the final step uses the fact that $n/(n - i) \le \epsilon^{-1} \le T^{o(1)}$.
    \item $|A_1| > \sqrt{T}$ and $|A_2| \le \sqrt{T}$, but $A = A_1$: For this to happen, $h_1(u)$ must be in the final $\sqrt{T}$ positions of $A$. Therefore, the contribution of this case to \eqref{eq:ab-bound} is at most
    \begin{align*} &
    \sum_{A \in \mathcal{L}} |A| \cdot \Pr[h_1(u) \text{ in final }\sqrt{T}\text{ positions of }A] \cdot \frac{n}{n - i} \\
    & \le \sum_{A \in \mathcal{L}} |A| \cdot \frac{\sqrt{T}}{n} \cdot \frac{n}{n - i}.
    \end{align*}
    Recalling by Lemma \ref{lem:long-runs-small-volume} that $\sum_{A \in \mathcal{L}} |A| \le O(n / T^{1 - o(1)})$ (with high probability), the sum above evaluates to at most
    $$O(n / T^{1 - o(1)}) \cdot \frac{\sqrt{T}}{n} \cdot \frac{n}{n - i} = O(1/T^{1/4}),$$
    where the final step uses the fact that $n/(n - i) \le \epsilon^{-1} \le T^{o(1)}$.
\end{enumerate}
Combining these cases, and analyzing the case of $A = A_2$ analogously, we obtain that
\begin{equation}\E[|A| \cdot |B| \cdot \mathbf{1}[|A| \ge T]] \le O(\E[\Phi_L] / T^{1/8}) + O(1/T^{1/4}).\label{eq:foob1}\end{equation}

Finally, we bound
$$\E[|A| \cdot |B| \cdot \mathbf{1}[|B| \ge T, |A| < T]].$$
This is at most
\begin{align*}
\E[(|A_1| + |A_2|) \cdot |B| \cdot \mathbf{1}[|B| \ge T, |A| < T]]
& = 2 \cdot \E[|A_1| \cdot |B| \cdot \mathbf{1}[|B| \ge T, |A| < T]] \\
& \le 2\E\left[\sum_{R \in \mathcal{R} \setminus \mathcal{L}} |R| \cdot \Pr[A_1 = R] \cdot \E[B \cdot \mathbf{1}[B \in \mathcal{L}] \mid A_1 = R]\right] \\
& \le 2\E\left[\sum_{R \in \mathcal{R} \setminus \mathcal{L}} \frac{|R|^2}{n} \cdot \E[B \cdot \mathbf{1}[B \in \mathcal{L}] \mid A_1 = R]\right] \\
& \le 2 \E\left[\sum_{R \in \mathcal{R} \setminus \mathcal{L}} \frac{|R|^2}{n} \cdot \sum_{B \in \mathcal{L}} \frac{1}{n - i}\right] \\
& = 2 \E[\Phi_T] \cdot |\mathcal{L}| \cdot \frac{1}{n - i} \\
& \le 2 \E[\Phi_T] \cdot \frac{n / T}{n - i} \\
& \le \E[\Phi_T] / T^{1/2}. \tag{since $n/(n - i) \le \epsilon^{-1} \le T^{o(1)}$}
\end{align*}
Combining this with \eqref{eq:foob1}, and plugging these into our original expression for $(\Phi_L' - \Phi_L) \cdot n$, we can conclude that
$$(\E[\Phi_L'] - \E[\Phi_L]) \cdot n \le O(\E[\Phi_L] / T^{1/8}) + O(1/T^{1/4}) + O(\E[\Phi_T] / T^{1/2}) + \E[\Psi(i)],$$
as desired.
\end{proof}

Although bounding each individual $\Psi(i)$ term is difficult, we can argue that the sum of the $\Psi(i)$ terms is directly related to $\Phi_T$.
\begin{lemma}
If $\Phi_T$ is the value of $\Phi_T$ prior to the $(1 - \epsilon)n$-th insertion, then $\sum_{i = 1}^{(1 - \epsilon)n - 1} \Psi(i) \le O(n \cdot \Phi_T)$.
\label{lem:Psi}
\end{lemma}
\begin{proof}
If $\Psi(i) \neq 0$, then the $i$-th insertion must combine runs $A^{(i)}$ and $B^{(i)}$ of length $< T$ into a run $C^{(i)}$ of length $\ge T$. The runs $\{C^{(i)} \mid \Psi(i) \neq 0\}$ are disjoint from one-another. So, if we define $\mathcal{L}$ to be the set of runs of length $\ge T$ after the $(1 - \epsilon)n$-th insertion, then
$$\sum_{i \le (1 - \epsilon)n \mid \Psi(i) \neq 0} |C^{(i)}| \le \sum_{R \in \mathcal{L}} |R|.$$
Notice, however, that
$$\Phi_T \ge T \sum_{R \in \mathcal{L}} |R| / n,$$
meaning that
$$\Phi_T \ge T \sum_{i  \le (1- \epsilon)n \mid \Psi(i) \neq 0} |C^{(i)}| / n = \Theta\left( \sum_{i  \le (1- \epsilon)n \mid \Psi(i) \neq 0} |C^{(i)}|^2 / n\right),$$
where the final step uses that, if $\Psi(i) \neq 0$, then $|C^{(i)}| = \Theta(T)$.
\end{proof}

Finally, we can put the pieces together to obtain a bound on the final value of $\Phi_L$. This bound basically reduces the task of analyzing $\E[\Phi_L]$ to the task of analyzing $\E[\Phi_T]$.

\begin{lemma}
Let $\Phi_L$ and $\Phi_T$ be the long potential and the truncated potential, respectively, right before the $(1 - \epsilon)n$-th insertion. Then,
$$\E[\Phi_L] \le O(\E[\Phi_T] + 1/T^{1/8}).$$
\label{lem:phiL}
\end{lemma}
\begin{proof}
Let $\Phi_T$ be the truncated potential right before the $(1 - \epsilon)n$-th insertion. By Lemma \ref{lem:foob1}, we have
\begin{align*}
    n \cdot \E[\Phi_L] & \le \sum_{i = 1}^{(1 - \epsilon)n - 1} O(\E[\Phi_L] / T^{1/8}) + O(1/T^{1/4}) + O(\E[\Phi_T] / T^{1/2}) + \E[\Psi(i)]  \\
    & \le  O(n / T^{1/4} + n \cdot \E[\Phi_T] + n \cdot \E[\Phi_L] / T^{1/8}) + \sum_{i = 1}^{(1- \epsilon)n-1} \Psi(i) \\
    & \le O(n / T^{1/4} + n \cdot \E[\Phi_T] + n \cdot \E[\Phi_L] / T^{1/8}). \tag{by Lemma \ref{lem:Psi}}
\end{align*}
Rearranging gives
$$(1 - O(T^{-1/8})) \cdot \E[\Phi_L] \le O(n \cdot \E[\Phi_T] + n / T^{1/4}),$$
which implies that
$$\E[\Phi_L] \le O(\E[\Phi_T] + 1/T^{1/8}),$$
as desired.
\end{proof}

Finally, we can define the full potential
$$\Phi = \sum_{R \in \mathcal{R}} |R|^2/n,$$
Note that $\Phi \le \Phi_T + \Phi_L$. Therefore, by Lemmas \ref{lem:phiT} and \ref{lem:phiL}, we have that
$$\E[\Phi] \le O(\E[\Phi_T] + 1/T^{1/8}) = O(\epsilon^{-(2 - \Omega(1))}).$$
Thus, we have proven the following key technical lemma:
\begin{lemma}
Immediately before the $(1 - \epsilon)n$-th insertion, the expected value of $\Phi$ is at most $O(\epsilon^{-(2 - \Omega(1))})$.
\label{lem:phi}
\end{lemma}

As an immediate corollary, we have the following bound on expected insertion time.

\begin{lemma}
The $(1- \epsilon)n$-th insertion takes expected time $O(\epsilon^{-(2 - \Omega(1))})$.
\label{lem:insertiontimebound}
\end{lemma}
\begin{proof}
Let $u$ be the element being inserted.
Let $A_1$ and $A_2$ be the runs containing $h_1(u)$ and $h_2(u)$, respectively, prior to the insertion. The expected insertion time is at most $\E[|A_1| + |A_2|] = O(\E[|A_1|])$. But
$$\E[|A_1|] = \sum_{R \in \mathcal{R}} |R| \cdot \Pr[A_1 = R] = \sum_{R \in \mathcal{R}} |R| \cdot \frac{|R|}{n} = \E[\Phi] = O(\epsilon^{-(2 - \Omega(1))}),$$
so the expected insertion time is at most $O(\epsilon^{-(2 - \Omega(1))})$.
\end{proof}

\paragraph{Bounding query time. }
Finally, we can now bound the expected query time. This argument is a bit more subtle, but once again follows from our bound on $\Phi$.

We begin by giving a simple characterization of how the query time for $u$ evolves over time.
\begin{lemma}
Consider an element $u$ with hashes $h_1(u)$ and $h_2(u)$ that may or may not be in the hash table. The time to query $u$ is at most the total number of times that insertions $u'$ select a run that contains either $h_1(u)$ or $h_2(u)$.
\label{clm:query-time-characterization}
\end{lemma}
\begin{proof}
Define $X_i$ to be the time to query $u$ prior to the $i$-th insertion. Let $Y_i$ be the indicator random variable for the event that the $i$-th insertion $u_i$ uses a run that contains either $h_1(u)$ or $h_2(u)$. Then, $X_0 = 0$, and we claim that $X_{i + 1} \le X_i + Y_i$. Indeed, if $Y_i = 0$, then the value of $X_i$ is clearly unchanged, and if $Y_i = 1$, then the query time may change. Notice, however, that during any given insertion, no element's query time increases by more than $1$: indeed, because of how we order elements, we have for every $j \in [n]$ that the maximum position to contain an element with hash $\le j$ increases by at most one (for $j$ close to $n$, this ``maximum position'' is permitted to wrap around the end of the array). Thus, if $Y_i = 1$, then the query time for $u$ increases by at most 1, implying that $X_{i + 1} \le X_i + Y_i$. Hence $X_{i + 1} \le \sum_{j = 0}^i Y_j$, as desired.
\end{proof}

Next, we show that for elements $u$ \emph{not} in the hash table, there is a simple upper bound on the probability of a given insertion increasing $u$'s query time.
\begin{lemma}
    Consider an element $u$ with hashes $h_1(u)$ and $h_2(u)$ that is \emph{not} in the hash table. Let $R$ be the run containing $h_1(u)$ prior to the $(1 - \epsilon)n$-th insertion. Then, so long as the $i$-th insertion does not insert $u$, the probability of the $i$-th insertion having a hash in $R$ is $O(\epsilon^{-(2 - \Omega(1))} / n)$.
    \label{lem:negrunbound}
\end{lemma}
\begin{proof}
The expected length of $R$ is $\E[\sum_{R \in \mathcal{R}} |R|^2/ n] = \E[\Phi] = O(\epsilon^{-(2 - \Omega(1))})$, where the final step uses Lemma \ref{lem:phi}. The probability of the $i$-th insertion having a hash in $R$ is therefore $O(\epsilon^{-(2 - \Omega(1))} / n)$, as desired.
\end{proof}

The more interesting case is the one where $u$ \emph{is} in the hash table. The fact that the run $R$ containing $u$ is known to contain $u$ means that we cannot treat $R$ as uniformly random from $\mathcal{R}$. Fortunately, we can leverage Lemma \ref{lem:lipshitz} to show that, nonetheless, the bound in Lemma \ref{lem:negrunbound} still holds.
\begin{lemma}
    Consider an element $u$ with hashes $h_1(u)$ and $h_2(u)$ that \emph{is} inserted into the hash table prior to the $(1 - \epsilon)n$-th insertion. Let $R$ be the run containing $h_1(u)$ prior to the $(1 - \epsilon)n$-th insertion. Then the probability of the $(1 - \epsilon)n$-th insertion having a hash in $R$ is $O(\epsilon^{-(2 - \Omega(1))} / n)$.
    \label{lem:posrunbound}
\end{lemma}
\begin{proof}
Let $W_1$ be a world in which the first $i$ insertions are the same, but with $u$ removed, and let $W_2$ be the (actual) world in which $u$ is present. By Lemma \ref{lem:lipshitz}, the run containing $h_1(u)$ in $W_2$ is a sub-interval of the concatenation $R_0 \circ R_1 \circ R_2$, where $R_1$ is the run containing $u$ in world $W_1$, and $R_0, R_2$ are the runs that appear directly before and after it. By Lemma \ref{lem:negrunbound}, the probability of the $(1 - \epsilon)n$-th insertion having a hash in $R_1$ is $O(\epsilon^{-(2 - \Omega(1))} / n)$. Since $R_0, R_2$ are uniformly random from $\mathcal{R} \setminus \{R_1\}$, we have by the same reasoning as in Lemma \ref{lem:boundB} that $\E[|R_0|]$ and $\E[|R_2|]$ are $O(\epsilon^{-1})$, meaning that the probability of the $(1 - \epsilon)n$-th insertion having a hash in $R_0$ or $R_2$ is $O(\epsilon^{-1}/n) = O(\epsilon^{-(2 - \Omega(1))} / n)$. This completes the proof.
\end{proof}

Finally, we can put the pieces together to obtain a bound on the expected query time.
\begin{lemma}
Consider an element $u$ with hashes $h_1(u)$ and $h_2(u)$ that may or may not be in the hash table. Prior to the $(1 - \epsilon)n$-th insertion, the expected time to query $u$ is $O(\epsilon^{-(1 - \Omega(1))})$.
\label{lem:querytimebound}
\end{lemma}
\begin{proof}
By Lemmas \ref{lem:phi}, \ref{lem:negrunbound}, and \ref{lem:posrunbound}, we have that the expected query time is at most
$$O(1) + \sum_{i = 1}^{(1 - \epsilon)n - 1} O\left(\frac{\left(\frac{n}{n - i}\right)^{2 - \Omega(1)}}{n}\right) = O\left(\int_{\delta = \epsilon}^{1} \delta^{-(2 - \Omega(1))} d\delta\right) = O(\epsilon^{-(1 - \Omega(1))}).$$
\end{proof}

Combining Lemmas \ref{lem:querytimebound} and \ref{lem:insertiontimebound}, we have proven Theorem \ref{thm:greedy}.

\section{A Two-Choice Algorithm With $O(\log\epsilon^{-1})$ Queries}\label{sec:nongreedy}

In this section, we show how to design a \emph{non-greedy} two-choice protocol that reduces the expected query time not just to $\epsilon^{-(1 - \Omega(1))}$, as in Section \ref{sec:greedy}, but all the way down to $O(\log \epsilon^{-1})$, while also achieving an amortized expected insertion time of $O(\epsilon^{-1})$.

\thmNonGreedy*

We remark that, in addition to storing the hash table itself, the implementation that we use to prove Theorem \ref{thm:non-greedy} will also store $2n$ bits of metadata. As discussed in Section \ref{sec:prelims}, this metadata requirement can, in principle, be removed relatively straightforwardly by simply encoding the metadata into the relative order of elements (i.e., using something that is not quite ordered linear probing).

\paragraph{Section-level conventions. }As a convention, throughout this section, we will assume (except when stated otherwise) that we are analyzing an insertion/query at a load factor of $1 - \epsilon$ (i.e., the load factor prior to the operation is $ 1- \epsilon$), and that $\epsilon^{-1} = n^{o(1)}$. Specifically, when discussing a sequence of insertions, we will sometimes use $\epsilon$ as a live variable that adopts a different value for each insertion.

Throughout the section, we will find it helpful to break time into $O(\log \epsilon^{-1})$ \defn{phases}, where phase $i$ begins once the load factor reaches $1 - t_i$, with $t_i = 0.9^i$ for $i \in [0, \lceil \log_{0.9} \epsilon \rceil)$, and $t_i = \epsilon$ for $i = \lceil \log_{0.9} \epsilon \rceil$. (For convenience, we also define $t_{\infty} = 0$.) Note that the first phase is phase $0$, and that the final phase $\lceil \log_{0.9} \epsilon \rceil$ contains no insertions.

Recall that, for elements $x$ that are in the hash table, we use $h(x)$ to refer to whichever hash function $h_1(x)$ or $h_2(x)$ the element chose to use during insertion. A two-choice ordered linear probing hash table will always keep the elements in increasing order of $h$. As an abuse of notation in this section, when discussing $h$ as a hash function, we will treat it as an integer in $[n]$, but when discussing the ordering of elements, we will treat $h$ as a real number in $[0, n]$, so that the ordering is fully specified by the hash function.

\subsection{Warmup: A Solution Using Two-Way Probing}\label{sec:nongreedywarmup}

To motivate our algorithm for Theorem \ref{thm:non-greedy}, we begin with a warmup algorithm that makes three simplifications:
\begin{itemize}
    \item \textbf{Slot-Usage Flexibility: }We allow the algorithm to decide which slots to put elements in, where the only constraint is that elements appear from left to right in increasing order of hash. There is no requirement as to where free slots are, and an element $x$ may even be in a position to the left of $h(x)$.
    \item \textbf{Two-Way Probing: }To ensure that queries find elements $x$ that are to the left of $h(x)$, the query algorithm probes both to the left and to the right.
    \item \textbf{Simplified Query Time: }When performing a query, we do not count the time spent probing free slots, only the time spent probing occupied slots.\footnote{We remark that this final simplification is not strictly necessary, even for the warmup algorithm that we will present next. We include the assumption only to streamline the warmup, and since the assumption will naturally disappear in the full construction.}
\end{itemize}
We call a solution that makes these simplifications a \defn{weak} two-choice linear-probing hash table.

In this section, we will show how to construct a (surprisingly elegant!) weak solution that matches the guarantees promised in Theorem \ref{thm:non-greedy}. We will then show in Section \ref{sec:nongreedyfull} how to transform our weak solution into a full solution that proves Theorem \ref{thm:non-greedy}.

\begin{proposition}
    Let $\epsilon^{-1} = n^{o(1)}$. There is an implementation of weak ordered two-choice linear probing that, at load factor $1 - \epsilon$, supports queries with expected simplified query time $O(\log \epsilon^{-1})$ and that supports insertions with amortized expected time $O(\epsilon^{-1})$. In addition to storing the hash table itself, the implementation uses $n$ bits of metadata.
    \label{prop:logwarmup}
\end{proposition}

\paragraph{The Basic Algorithm. }The algorithm maintains, at any given moment, a partition $P$ of the array into disjoint blocks, where each block $B$ is guaranteed to have a number of free slots $\texttt{fr}(B)$ satisfying
\begin{equation}
\texttt{fr}(B) \in [0.5 \cdot \epsilon |B|, 1.5  \cdot \epsilon |B|].
\label{blockrestriction}
\end{equation}
We will come back in a moment to how the algorithm generates the partition $P$, and focus for now instead on how insertions are performed.

To insert an element $x$, where the load factor prior to the insertion is $1 - \epsilon$, the algorithm determines which two blocks $B_1$ and $B_2$ contain $h_1(x)$ and $h_2(x)$, respectively. (Note that $B_1$ may equal $B_2$.) The algorithm then selects hash $h_1$ with probability
$$\frac{1}{2} + \frac{\texttt{fr}(B_1) / |B_1| - \texttt{fr}(B_2) / |B_2|}{2\epsilon},$$
and selects $h_2$ otherwise. Note that this probability is, by \eqref{blockrestriction}, guaranteed to reside in $[0, 1]$.

If the algorithm selects $h_i$ for some $i \in \{1, 2\}$, then the algorithm selects a uniformly random free slot $s \in B_i$. Finally, the algorithm inserts $x$ and rearranges the elements in $B_i$ so that slot $s$ becomes occupied, so that the occupied/free status of other slots is unchanged, and so that the elements continue to appear in sorted order of hash.

The key design principle behind this protocol is to maintain the following invariant: that the set of free slots, at any given moment, is \emph{uniformly random}.

\begin{lemma}
Consider an insertion at load factor $1 - \epsilon$. Suppose that, prior to the insertion, the set of free slots is uniformly random out of all $\epsilon n$-size subsets of $[n]$. Then, after the insertion, the set of free slots is uniformly random out of all $(\epsilon n - 1)$-size subsets of $[n]$.
\label{lem:randomfreeslots}
\end{lemma}
\begin{proof}
It suffices to show that each free slot $s$, prior to the insertion, is equally likely to be selected by the insertion. For this, in turn, it suffices to show that the probability of a given block $B \in P$ being selected is equal to
$$\frac{\texttt{fr}(B)}{\epsilon n}.$$
Notice, however, that the probability of block $B$ being chosen is exactly
$$2 \sum_{B' \in P} \Pr[B_1 = B\text{ and } B_2 = B'] \cdot \left(\frac{1}{2} + \frac{\texttt{fr}(B) / |B| - \texttt{fr}(B') / |B'|}{2\epsilon}\right),$$
which expands to
\begin{align*}
    & 2 \sum_{B' \in P} \frac{|B| \cdot |B'|}{n^2} \cdot \left(\frac{1}{2} + \frac{\texttt{fr}(B) / |B| - \texttt{fr}(B') / |B'|}{2\epsilon}\right) \\
    & = 2 \frac{|B|}{n^2} \sum_{B' \in P}  \cdot \left(\frac{|B
'|}{2} + \frac{\texttt{fr}(B) |B'| / |B| - \texttt{fr}(B')}{2\epsilon}\right) \\
    & = 2 \frac{|B|}{n^2} \left( \frac{n}{2} + \frac{n \texttt{fr}(B)  / |B| - \epsilon n}{2\epsilon}\right) \\
    & = |B|/n + \frac{\texttt{fr}(B)  - \epsilon |B|}{\epsilon n} \tag{since $\sum |B'| = n$ and $\sum \texttt{fr}(B') = \epsilon n$} \\
    & = \frac{\texttt{fr}(B)}{\epsilon n},
 \end{align*}
 as desired.
\end{proof}

\paragraph{Determining the partition $P$.} To complete the specification of the algorithm, we must also describe how to determine the partition $P$ used at any given moment. Note that we can store $P$ using $n$ bits of metadata (hence the $n$ bits in Proposition \ref{prop:logwarmup}), where the $i$-th bit is $1$ iff the $i$-th slot is the first slot in some block.

Construct a binary tree $\mathcal{T}$ on the slots of the array, where the root node (level $0$) represents the entire array, and the nodes at level $i \ge 1$ are given by intervals of the form $[\lceil jn/2^i \rceil, \lfloor (j + 1)n/2^i\rfloor]$, for $j \in \{0, 1, 2, \ldots, 2^i - 1\}$.  This is to say that the nodes in level $i$ represent disjoint subarrays of size $n / 2^i \pm 1$, and where the children $x$ and $y$ of a given (internal) node $z$ satisfy $x \cup y = z$. If a node has only one child, because both the node and its child represent an interval of size $1$, then remove the child and consider the node to be a leaf.

We will maintain the invariant that $P$ partitions the array into nodes in the tree $\mathcal{T}$.
Initially, the partition makes each leaf its own block. Then, over time, blocks are merged according to the following two \defn{merge policies}:
\begin{enumerate}
    \item Consider an insertion at load factor $1 - \epsilon$ that places an element in some block $B \in P$. If the insertion decreases $\texttt{fr}(B)/|B|$ below $0.5 \cdot \epsilon$, then the parent $p(B) \in \mathcal{T}$ of $B$ gets added to $P$, and all strict descendents of $p(B)$ get removed. This procedure then repeats recursively on $p(B)$ as necessary.
    \item Every time that the load factor $1 - \epsilon$ crosses a threshold of the form $1 - t_i$ (i.e., at the start of each phase), we update the blocks as follows. For each block $B \in P$, we check if $\texttt{fr}(B)/|B| > 1.1 \epsilon$; if so, then the parent $p(B) \in \mathcal{T}$ of $B$ gets added to $P$, and all strict descendents of $p(B)$ get removed. This procedure then repeats recursively on $p(B)$ as necessary.
\end{enumerate}

\paragraph{Proving the correctness of $P$. }We now argue that the above policies result in a valid partition $P$.
\begin{lemma}
Consider a point in time at which the load factor is $1 - \epsilon$. Then each block $B \in P$ satisfies $\texttt{fr}(B) / |B| \in [0.5\epsilon, 1.5 \epsilon]$.
\end{lemma}
\begin{proof}
We first argue that the merge policies never allow the quantity $N(B) := \texttt{fr}(B)/|B| - 0.5\epsilon$ to become negative for any block $B$. Note that, since $\epsilon$ only decreases over time, changes to $\epsilon$ cannot make $N(B)$ become negative. Moreover, merging a collection of blocks $B_1, B_2, \ldots$ for which $N(B_1), N(B_2), \ldots \ge 0$ into a new block $B$ necessarily results in $N(B) \ge 0$, so merges performed by merge policy (2) also will not make $N(B)$ negative for any block. Thus the only way for $N(B)$ to become negative is if an insertion \emph{into} that block causes $N(B)$ to become negative -- but, in that case, merge policy (1) will merge blocks appropriately to avoid there being any blocks $B' \in P$ with negative $N(B')$.

Next we argue that the merge policies never allow the quantity $M(B) := 1.5 \epsilon - \texttt{fr}(B)/|B|$ to become negative. Notice that merging a collection of blocks $B_1, B_2, \ldots$ for which $M(B_1), M(B_2), \ldots \ge 0$ into a new block $B$ necessarily results in $M(B) \ge \min(M(B_1), M(B_2), \ldots)$,  so merges performed by merge policy (1) do not decrease the quantity $\min_{B \in P} M(B)$. Furthermore, each insertion impacts $M(B)$, for a given block $B$, by possibly increasing $-\texttt{fr}(B)/|B|$ and by decreasing $1.5\epsilon$ by $1.5/n$. Thus, the total decrease to $\min_{B \in P} M(B)$ per insertion is at most $1.5/n$. On the other hand, merge policy (2) guarantees that, when the load factor crosses a threshold of the form $1 - t_i$, where $t_i$ is a power of $0.9$, we have
$$\min_{B \in P} M(B) \ge 1.5t_i - 1.1t_i \ge 0.4 t_i.$$ Since there are at most $0.1 t_i$ insertions between load factors $1 - t_i$ and $1 - t_{i + 1}$, it follows that, during those insertions, we have $\min_{B \in P} M(B) \ge 0.4 t_i - 1.5 \cdot (0.1 t_i) > 0$. This completes the proof that $M(B)$ remains positive for all blocks $B \in P$ at all times, which completes the proof of the lemma.
\end{proof}

\paragraph{Proving time efficiency. }
To analyze time efficiency, we begin with two simple claims. The first gives a necessary condition for a given node $u \in \mathcal{T}$ to be a block at the start of some phase $i$.
\begin{claim}
    Consider some internal node $u$ in $\mathcal{T}$. Let $a$ and $b$ be the children of $u$. In order for $B = u$ to occur at the start of phase $i$, there must exist some phase $j \le i$ such that, at the beginning of phase $j$, at least one of the children $x \in \{a, b\}$ had a number of free slots outside of the range
    \begin{equation*}
    [0.9 t_j |x|, 1.1 t_j|x|].
    \end{equation*}
    \label{clm:corrupt}
\end{claim}
\begin{proof}
This follows from straightforward casework. In order for $B = u$ to occur at the start of phase $i$, one of two things must have happened:
\begin{enumerate}
    \item Merge policy (1) added $u$ to $P$. Suppose this occurred at a load factor $1 - \delta$ with $\delta \in [t_{j - 1}, t_{j}]$, for some $j \le i$. At that point in time, at least one of the children  $x \in \{a, b\}$ must have had fewer than $0.5 \delta |x|$ free slots. It follows that, at the start of phase $j$, the number of free slots in $x$ must have been less than
    $$0.5 \delta |x| \le 0.5 t_{j} |x| / 0.9 \le 0.9 t_{j} |x|.$$
    \item Merge policy (2) added $u$ to $P$. This would have to occur at the start of some phase $j \le i$, at which point at least one of the children $x \in \{a, b\}$ must have had more than
    $1.1 t_j |x|$
    free slots.
\end{enumerate}
\end{proof}

The next claim produces a simple coupling that will allow us to -- later in our analysis -- handle situations where one insertion's randomness has already been spoiled to us.
\begin{claim}
Let $m < n$ and $r \in [m]$. Define $X = x_1, x_2, \ldots, x_m$, where $x_i$ is uniformly random out of $[n] \setminus \{x_1, x_2, \ldots, x_{i - 1}\}$. Define $Y = y_1, y_2, \ldots, y_{m + 1}$, where for $i \neq r$, $y_i$ is uniformly random out of $[n] \setminus \{y_1, y_2, \ldots, y_{i - 1}\}$, but for $i = r$, $y_i$ is chosen by an adversary who knows $y_1, y_2, \ldots y_{i - 1}$. Define $X_j = \{x_i \mid i \le j\}$ and $Y_j = \{y_i \mid i \le j\}$. There exists a coupling $(X, Y)$ such that, for all $j \in [m]$, $|X_j \Delta Y_j| \le 2$.
\label{clm:coupling}
\end{claim}
\begin{proof}
Set $x_i = y_i$ for $i < r$, and allow the adversary to select $y_r$. Then for $i \in [r, m - 1]$, select $x_i$ and $y_{i + 1}$ as follows, so that we can maintain the invariant $|X_{i - 1} \Delta Y_{i}| \le 1$. Select a uniformly random $x_{i} \in [n]$, and set $y_{i + 1}$ to be $x_i$ iff $x_i \not\in Y_i$. If $x_i \in Y_i$ (i.e., $x_i$ is the unique element in $Y_i \setminus X_{i - 1}$), then set $y_{i + 1}$ to be uniformly random out of $[n] \setminus Y_{i}$. This maintains the invariant that $|X_{i - 1} \Delta Y_{i}| \le 1$, while selecting both $x_{i}$ and $y_{i + 1}$ according to the correct distributions. Finally, having constructed $X_{m - 1}$ and $Y_m$, with $|X_{m - 1} \Delta Y_m| \le 1$, we select $x_m$ uniformly at random from $[n]$. By construction, $|X_j \Delta Y_j| \le 2$ for all $j \in [m]$, as desired.
\end{proof}

To bound insertion time (and also simplified query time), the main task is to bound the size of the block $B \in \mathcal{T}$ containing a given slot $s$. Note, however, that the slot $s$ is not necessarily independent of the hash table state -- rather, $s$ may be a hash of some known element $v$ (e.g., the element we are inserting or querying) that may or may not be in the hash table. The following lemma bounds the size of $B$, even if $s$ is given by a hash of a known element.
\begin{lemma}
Consider a point in time at which the load factor is $1 - \epsilon$. Let $s \in [n]$, let $v$ be an element (which may be in the hash table), and condition on $h_\ell(v) = s$ for some $\ell \in \{1, 2\}$. Let $B \in P$ be the block containing position $s$. Then, for all $k \ge 1$,
$$\Pr[|B| \ge \epsilon^{-1} k \mid h_\ell(v) = s] \le e^{-\Omega(k)}.$$
\label{lem:blockprob}
\end{lemma}
\begin{proof}
Throughout the proof, let $i = \lceil \log_{0.9} \epsilon \rceil$ so that $\epsilon = t_i$.
Consider some node $u$ in $\mathcal{T}$ that is an ancestor of the leaf corresponding to slot $s$. Let $a$ and $b$ be the children of $u$. By Claim \ref{clm:corrupt}, in order for $B = u$ to occur, there must exist some phase $j \le i$ such that, at the beginning of phase $j$, at least one of the children $x \in \{a, b\}$ had a number of free slots outside of the range
\begin{equation}
[0.9 t_j |x|, 1.1 t_j|x|].
\label{eq:corrupt}
\end{equation}

Next we consider the probability, for a given node $x$, of \eqref{eq:corrupt} occurring at the start of some phase $j$ (conditioned on $h_\ell(v) = s$, where $v$ and $s$ are as in the lemma statement). Let $S$ be the set of occupied slots at the start of phase $j$. Note that, because of the condition $h_\ell(v) = s$, $S$ is not (quite) uniformly random from $[n]$. Fortunately, by Claim \ref{clm:coupling}, we know that $S$ can be coupled to a uniformly random set $U \subseteq [n]$ of the same size, satisfying $|U \Delta S| \le 2$. In order for \eqref{eq:corrupt} to occur, the random set $U$ would need to satisfy
$$|([n] \setminus U) \cap x| \notin [0.9 t_j|x| + 1, 1.1 t_j|x| - 1].$$
Since $\E[|([n] \setminus U) \cap x|] = \epsilon |x| = t_j |x|$, and since $[n] \setminus U$ is a random subset of $[n]$, we can apply a Chernoff bound for sampling without replacement to conclude that \eqref{eq:corrupt} occurs with probability at most
\begin{equation}
e^{-\Omega(t_j |x|)}.
\label{eq:corruptprob}
\end{equation}

Since \eqref{eq:corruptprob} bounds the probability of \eqref{eq:corrupt} occurring during a given phase $j$, the probability, for a given node $x$, of \eqref{eq:corrupt} occurring at the start of \emph{any} phase $j \le i$ is at most
$$\sum_{j = 0}^i e^{-\Omega(t_j |x|)} = 2^{-\Omega(t_i |x|)}.$$
We can therefore conclude that, for an internal node $u \in \mathcal{T}$, the probability that \eqref{eq:corrupt} occurs to either of $u$'s children at the start of any phase $j \le i$ is at most $2^{-\Omega(t_i |u|)}.$ It follows by \eqref{eq:corrupt} that the probability of $B = u$, for any given node $u$, is also at most $2^{-\Omega(t_i |u|)}.$ Applying this bound to each (strict) ancestor of the leaf containing slot $s$ gives
$$\Pr[|B| \ge j \mid h_\ell(v) = s] \le 2^{-\Omega(j t_i)}.$$
Since $t_i = \epsilon$, the desired tail bound follows.
\end{proof}

With the help of Lemma \ref{lem:blockprob}, we can now bound amortized expected insertion time.
\begin{lemma}
The amortized expected time per insertion is at most $O(\epsilon^{-1})$.
\label{lem:couplinginsertion}
\end{lemma}
\begin{proof}
Consider the insertion of an element $v$ at load factor $1 - \epsilon$.
    The reason that the insertion time is amortized is that we must consider the time spent implementing merge policy (2) between phases. This contributes $O(n)$ time between phases $i$ and $i + 1$, for $i + 1 < \log_{0.9} \epsilon$, which can be paid for at a cost of $O(t_i^{-1})$ to each insertion in phase $i$. This contributes $O(\epsilon^{-1})$ to the amortized insertion time of $v$.

Besides the time spent implementing merge policy (2), the remaining time to insert $v$ is at most $O(|B_1| + |B_2|)$, where $B_1, B_2 \in \mathcal{T}$ are the blocks containing slots $h_1(v)$ and $h_2(v)$, respectively, after the insertion is complete. By Lemma \ref{lem:blockprob}, we know that $\E[|B_1|]$ and $\E[|B_2|]$ are each at most $O(\epsilon^{-1})$. This completes the proof.
\end{proof}

Next, we turn to the task of bounding (simplified) query time. Suppose the hash table has been filled to $1 - \epsilon$ load factor. For an element $v$ (not necessarily in the hash table), and for $i \in [2]$, define $A^i_v$ (resp.~$B^i_v$) to be the number of elements $y$ to the left (resp.~right) of $h_i(v)$ but with $h(y) \ge h_i(v)$ (resp.~$h(y) \le h_i(v)$). By symmetry, all of $A^1_v, B^1_v, A^2_v, B^2_v$ are identically distributed. The following claim demonstrates how to bound $A^1_v$.

\begin{claim}
Let $v$ be an element. For $j \in [(1 - \epsilon)n]$, define the indicator random variable $I_j$ to be $1$ iff the $j$-th insertion fills a free slot in the block $B \in \mathcal{T}$ that (prior to the insertion) contains slot $h_1(v)$. Then, $A^1_v \le \sum_{j = 1}^{(1 - \epsilon)n} I_j$.
\label{clm:a1v}
\end{claim}
\begin{proof}
First note that, by construction, the elements residing in a given block $B$ all have hashes within $B$. This is because, when an insertion is performed within some block, the rearrangement of elements that occurs only moves elements around \emph{within} that block.

Thus the elements $y$ to the left of $h_1(v)$ but with hashes $h(y) \ge h_1(v)$ are all in the block containing $h_1(v)$. Insertions that fill free slots outside of the block have no effect on the number of such $y$. Insertions that fill a free slot inside the block can increase by the number of such $y$ by at most $1$. Thus the claim follows.
\end{proof}

Using Claim \ref{clm:a1v}, we can bound $\E[A^1_v]$ as follows:
\begin{claim}
$\E[A^1_v] \le O(\log \epsilon^{-1})$.
\label{clm:ea1v}
\end{claim}
\begin{proof}
Defining the $I_j$s as in Claim \ref{clm:a1v}, we have
$$A^1_v \le \sum_{j = 1}^{(1 - \epsilon)n} I_j.$$
Now consider the $j$-th insertion for some $j \in [(1 - \epsilon)n]$, and define $\delta_j = (n - j) /n$. If the element $y$ being inserted is not $v$, and if we define the block $B \in \mathcal{T}$ to be the block containing slot $h_1(y)$ prior to the insertion, then
$$\Pr[I_j = 1] \le 2|B| /n.$$
By Lemma \ref{lem:blockprob}, we know that $\E[|B|] \le O(\delta_j^{-1})$, meaning that
$$\E[I_j] \le O(\delta_j^{-1}/n).$$
It follows that
$$\E[A^1_v] \le 1 + \sum_{j = 1}^{(1 - \epsilon)n} O(\delta_j^{-1} /n) = 1 + \sum_{j = 1}^{(1 - \epsilon)n} O((n - j)^{-1}) = 1 + O(\log \epsilon^{-1}).$$
\end{proof}

Using the symmetry between $A^1_v, A^2_v, B^1_v, B^2_v$, we can bound simplified query time as follows:
\begin{lemma}
    Suppose the hash table is filled to load factor $1 - \epsilon$. Then, the expected simplified query time to query an element $v$ is at most $O(\log \epsilon^{-1})$.
\end{lemma}
\begin{proof}
Since $A^1_v, A^2_v, B^1_v, B^2_v$ are identically distributed, we know from Claim \ref{clm:ea1v} that
$$\E[A^1_v + A^2_v + B^1_v + B^2_v] = O(\log \epsilon^{-1}).$$
It follows that the expectation of the simplified query time to query $v$ is at most $O(\log \epsilon^{-1})$, as desired.
\end{proof}

This completes the proof of Proposition \ref{prop:logwarmup}.

\subsection{Removing the Simplifying Assumptions}\label{sec:nongreedyfull}

To prove Theorem \ref{thm:non-greedy}, we will need to remove the simplifying assumptions that we made in the previous section. Our basic approach will be as follows: We will continue to use the construction from the previous section to determine our two-choice policy (i.e., which hash $h \in \{h_1, h_2\}$ each element uses). But, given this two-choice policy, we will then place the elements into the hash table using standard ordered linear probing, where each element is placed according to its chosen hash function.

Note that our new algorithm no longer maintains the invariant that the set of free slots is uniformly random from $[n]$. So that we can refer to the state that the hash table \emph{would have been in}, if we had used the algorithm in the previous section, we will refer to this state as the \defn{virtual state}. Note that, in order to implement the two-choice policy, we will need to keep track of which slots are free/occupied in the virtual state. We can do this with $n$ bits of metadata, bringing the total number of metadata bits from $n$ (as in Proposition \ref{prop:logwarmup}) to $2n$ (as in Theorem \ref{thm:non-greedy}).

The main obstacle to proving Theorem \ref{thm:non-greedy} is to show that the new ordering of the elements does not produce significantly worse query times than in the previous section, i.e., that the query time remains $O(\log \epsilon^{-1})$. Our next lemma gives a simple way of relating query times in the new hash table to the quantities $A^i_j$ and $B^i_j$ defined in the previous section.

\begin{claim}
    Consider a query to an element $v$ whose first hash $h_1(v)$ is given by some position $j$. For $a < j$, define $F_{j - a, j}$ to be the number of free slots in $[j - a, j]$ in the \emph{virtual} state. Then the time to query $v$ from $h_1(v) = j$ is at most
    $$O\left(1 + B^1_j + \max_{a \ge 1} \left(A^1_{j - a} - \epsilon a / 2\right) + \max_{a \ge 1} \left(\epsilon a / 2 - F_{j - a, j}\right)\right).$$
    \label{clm:querytimebreakdown}
\end{claim}
\begin{proof}
Let $R$ be the maximal interval of the form $[j - a, j + b]$, where all of the slots are occupied (no free slots) by elements $u$ with hashes $h(u) \le j$. By construction, all the elements $u$ in $R$ are exactly those with hashes $h(u) \in [j - a, j]$. The number of elements in $R$ is therefore at most
$$A^1_{j - a} + B^1_{j} + (a + 1) - F_{j - a, j},$$
and the time $\Theta(b)$ to search for $v$ starting from position $j = h_1(v)$ satisfies
$$b \le |R| - a \le 1 + A^1_{j - a} + B^1_{j} - F_{j - a, j}.$$
We can therefore bound $b$ by
$$1 + B^1_j + \max_{a \ge 1} \left(A^1_{j - a} - F_{j - a, j}\right).$$
This, in turn is at most
$$1 + B^1_j + \max_{a \ge 1} (A^1_{j - a} - \epsilon a / 2) + \max_{a \ge 1} (\epsilon a / 2 - F_{j - a, j}),$$
as desired.
\end{proof}

Our next step in bounding the query time is to bound the term involving $F_{j - a, j}$.
\begin{claim}
    Consider a query to an element $v$ whose first hash $h_1(v)$ is given by some position $j$. Then,
    $$\E\left[\max_{a \ge 1} (\epsilon a / 2 - F_{j - a, j})\right] \le O(1).$$
    \label{clm:free}
\end{claim}
\begin{proof}
By Claim \ref{clm:coupling}, it suffices to consider a table state in which the $\epsilon n$ slots that are free are uniformly random, and to show for such a table state that
$$\E\left[\max_{a \ge 1} (\epsilon a / 2 - F_{j - a, j})\right] \le O(1).$$
Notice, however, that
$$\E\left[\max_{a \ge 1} (\epsilon a / 2 - F_{j - a, j})\right] \le O(1) + \E\left[\max_{a \text{ a multiple of } \epsilon^{-1}} (\epsilon a / 2 - F_{j - a, j})\right].$$
By a Chernoff bound for sampling without replacement, since $\E[F_{j - a, j}] = \epsilon a$, we know that $\Pr[F_{j - a, j} \le \epsilon a / 2] \le e^{-\Omega(\epsilon a)}$. It follows that
$$\E[\max(0, \epsilon a / 2 - F_{j - a, j})] \le e^{-\Omega(\epsilon a)}.$$
Summing over all multiples $a$ of $\epsilon^{-1}$, we get that
\begin{align*}
&\E\left[\max_{a \text{ a multiple of } \epsilon^{-1}} (\epsilon a / 2 - F_{j - a, j})\right] \\
&\qquad \le \E\left[\sum_{a \text{ a multiple of } \epsilon^{-1}} \max(0, \epsilon a / 2 - F_{j - a, j})\right] \\
&\qquad \le O(1) + \sum_{a = 1}^{\infty} e^{-\Omega(\epsilon a)} \\
&\qquad = O(1),
\end{align*}
as desired.
\end{proof}

The most difficult technical result in the analysis (and, indeed, in the entire section), will be to prove the following concentration bound on $A^1_{j - a}$.
\begin{proposition}
    Suppose the load factor is $1 - \epsilon$. Consider a query to an element $v$ whose first hash $h_1(v)$ is given by some position $j$. Then, for any $a$, we have that $A^1_{j - a}$ is, with high probability in $n$, at most $O(\log \epsilon^{-1})$ plus a geometric random variable with mean $O(1)$.
    \label{prop:difficult}
\end{proposition}
We defer the proof of Proposition \ref{prop:difficult} to the end of the section. For now, we will use it to complete our analyses of both query and insertion time.

\begin{lemma}
The expected time to query an element at load factor $1 - \epsilon$ is $O(\log \epsilon^{-1})$.
\end{lemma}
\begin{proof}
Let $v$ be the element we are querying, and set $j = h_1(v)$.
If we define a sufficiently large positive constant $c$, we can deduce from Lemma \ref{prop:difficult} that
\begin{align*}
\E\left[\max_{a \ge 1} \left(A^1_{j - a} - \frac{\epsilon a}{2}\right)\right]
&\le O\left(\log \epsilon^{-1}\right) + \E\left[\max_{a \ge 1} \left(A^1_{j - a} - \frac{\epsilon a}{2} - c \log \epsilon^{-1}\right)\right] \\
&\le O\left(\log \epsilon^{-1}\right) + \E\left[\sum_{a \ge 1} \max\left(0, A^1_{j - a} - \frac{\epsilon a}{2} - c \log \epsilon^{-1}\right)\right] \\
&\le O\left(\log \epsilon^{-1}\right) + \E\left[\sum_{a \ge 1} \epsilon e^{-\Omega(\epsilon a / 2)}\right] \tag{by Lemma \ref{prop:difficult}} \\
&\le O\left(\log \epsilon^{-1}\right) + O(1) = O\left(\log \epsilon^{-1}\right).
\end{align*}

Combining this with Claim \ref{clm:a1v} and Claim \ref{clm:free}, we conclude that the expected time to query $v$ from $h_1(v) = j$ is at most
$$B^1_j + \max_{a \ge 1} (A^1_{j - a} - \epsilon a / 2) + \max_{a \ge 1} (\epsilon a / 2 - F_{j - a, j}) \le O(\log \epsilon^{-1}).$$
By symmetry between $h_1$ and $h_2$, it follows that the expected time to query $v$ overall is $O(\log \epsilon^{-1})$.
\end{proof}

Finally, as a corollary of our bound for query time, we can extract a bound on amortized expected insertion time.

\begin{lemma}
The amortized expected insertion time is $O(\epsilon^{-1})$.
\end{lemma}
\begin{proof}
Notice that the time to insert $v$ is at most $C_1(v)$, the time to insert $v$ in the virtual state; plus $C_2(v)$, the time to query $v$; plus $C_3(v)$, the number of elements that the insertion moves. We already know that $C_1(v)$ contributes $O(\epsilon^{-1})$ to the amortized expected insertion time, and that $C_2(v)$ contributes $O(\log \epsilon^{-1})$. To bound the contribution of $C_3(v)$, note that, for each of the $C_3(v)$ elements that the insertion of $v$ moves, the query time of that element gets increased. It follows that the sum, over all inserted elements $u$ of $C_3(u)$ is at most the sum of the query times of the elements, which has expected value $O(n \log \epsilon^{-1})$. Thus, the expected sum of the $C_3(v)$ costs \emph{over all insertions} is at most $O((1 - \epsilon)n (1 +\log \epsilon^{-1}))$. To pay for this, it suffices to charge an amortized cost of $O(1 + n/(n - j))$ to the $j$-th insertion, that way the total charge across the first $(1 - \epsilon)n$ insertions is
$$O\left(\sum_{j = 1}^{(1 - \epsilon)n} \left(1 + \frac{n}{n - j}\right)\right) = \Theta((1 - \epsilon)n (1 + \log \epsilon^{-1})).$$
At load factor $1 - \epsilon$, this charge is $O(\epsilon^{-1})$. Combining the contributions of $C_1(v)$, $C_2(v)$, and $C_3(v)$ results in an amortized expected insertion time of $O(\epsilon^{-1})$, as desired.
\end{proof}

\subsection{Proving Proposition \ref{prop:difficult}}

In this section, we prove Proposition \ref{prop:difficult}, restated below (with a slightly different choice of variables, and with the condition $\epsilon^{-1} = n^{o(1)}$ from Theorem \ref{thm:non-greedy} made explicit).

\begin{proposition}
    Suppose the hash table is at load factor $1 - \epsilon$, where $\epsilon^{-1} = n^{o(1)}$. Consider a query to an element $v$ whose first hash $h_1(v)$ is given by some position $q$. Then, for any position $u$, we have that $A^1_{u}$ is, with high probability in $n$, at most $O(\log \epsilon^{-1})$ plus a geometric random variable with mean $O(1)$.
    \label{prop:difficult-a1}
\end{proposition}

Because $A^1_u$ concerns only the algorithm from Section \ref{sec:nongreedywarmup}, we will refer only to the state of the array according to that algorithm (i.e., we will focus on what, in Section \ref{sec:nongreedywarmup}, would be considered the \emph{virtual} state of the array). We will also, to simplify exposition, introduce the terminology of saying that a slot is \defn{activated} when an insertion fills it.

By Claims \ref{clm:corrupt} and \ref{clm:a1v}, to prove Proposition \ref{prop:difficult-a1}, it suffices to prove the following result. (It may seem strange to inline so many definitions in a proposition, but it will see a good reason for it in a moment.)

\begin{proposition}
Let $u, v$ be elements, let $q \in [n]$, and suppose $\epsilon^{-1} = n^{o(1)}$. Say that a node $x$ \emph{becomes corrupt} in phase $i$ if, at the start of phase $i$ its number of free slots is outside the interval $[t_i \cdot |x| \cdot 0.9, t_i \cdot |x| \cdot 1.1]$, and if it is not corrupted in any phase $j < i$. Once a node $x$ is corrupted, it is said to be \emph{corrupt} in all phases $j \ge i$. Finally, say that a node $x$ is \emph{dead} in phase $i$ if at least one of its children is corrupt.

Let $u_0, u_1, \ldots$ denote the ancestors of $u$, where $u_0 = u$ and $u_{i + 1}$ is the parent of $u_i$. (Additionally, define $u_{-1} = \emptyset$.) For each $u_i$, define $D(u_i)$ to be the phase number at the start of which $u_i$ dies, if such a phase exists, and $\infty$ otherwise. Finally, define $A_{i, j}$ to be $0$ if $j = \infty$, and to otherwise be the number of free slots that activate in $u_i$ during phases $j - 1, j, j + 1, \ldots$.

Then,
 $$\E\left[\sum_{i \ge 0} A_{i, D(u_i)} \mid h(v) = q\right]$$ is, with high probability in $n$, at most $O(\log \epsilon^{-1})$ plus a geometric random variable with mean $O(1)$.
 \label{prop:strawman}
\end{proposition}

The condition $h(v) = q$ in Proposition \ref{prop:strawman} is, a priori, a bit of a pain to deal with. Fortunately, we can remove it using Claim \ref{clm:coupling}. Recall that Claim \ref{clm:coupling} tells us that, if we compare a world in which we condition on $h(v) = q$ to one in which we don't, the difference is at most one edit to the set of slots that are free at any given moment. It therefore sufices to prove the following modified version of Proposition \ref{prop:strawman}, with the condition $h(v) = q$ removed the two other modifications boldfaced:

\begin{proposition}
    Let $u, v$ be elements, let $q \in [n]$, and suppose $\epsilon^{-1} = n^{o(1)}$. Say that a node $x$ \emph{becomes corrupt} in phase $i$ if, at the start of phase $i$ its number of free slots is outside the interval $[t_i \cdot |x| \cdot 0.9 \mathbf{+1}, t_i \cdot |x| \cdot 1.1 \mathbf{-1}]$, and if it is not corrupted in any phase $j < i$. Once a node $x$ is corrupted, it is said to be \emph{corrupt} in all phases $j \ge i$. Finally, say that a node $x$ is \emph{dead} in phase $i$ if at least one of its children is corrupt.

Let $u_0, u_1, \ldots$ denote the ancestors of $u$, where $u_0 = u$ and $u_{i + 1}$ is the parent of $u_i$. (Additionally, define $u_{-1} = \emptyset$.) For each $u_i$, define $D(u_i)$ to be the phase number at the start of which $u_i$ dies, if such a phase exists, and $\infty$ otherwise. Finally, define $A_{i, j}$ to be $0$ if $j = \infty$, and to otherwise be \textbf{one plus} the number of free slots that activate in $u_i$ during phases $j - 1, j, j + 1, \ldots$.

Then,
 $$\E\left[\sum_{i \ge 0} A_{i, D(u_i)} \mid h(v) = q\right]$$ is, with high probability in $n$, at most $O(\log \epsilon^{-1})$ plus a geometric random variable with mean $O(1)$.
     \label{prop:geo}
    \end{proposition}

The rest of the section is devoted to proving Proposition \ref{prop:geo}. Note that, throughout, we will use the definitions of corrupt and of $A_{i, j}$ given in Proposition \ref{prop:geo} (not the ones given in Proposition \ref{prop:strawman}).

To prove Proposition \ref{prop:geo}, we begin with a few simple observations. We first observe that $A_{i, D(u_i)}$ is straightforwardly bounded by $O(t_{D(u_i)} \cdot |u_i|)$.

\begin{lemma}
For a given $i \ge 0$, $A_{i, D(u_i)} \le 4 t_{D(u_i)} \cdot |u_i| + 1$.
\label{lem:remainingslotsbound}
\end{lemma}
\begin{proof}
Because $u_i$ dies in phase $D(u_i)$, we know that in the beginning of phase $D(u_i) - 1$, $u_i$'s children were not yet corrupted. Therefore, the number of free slots in $u_i$ at the beginning of $D(u_i) - 1$ is at most $2 t_{D(u_i) - 1} \cdot |u_i| \le 4 t_{D(u_i)} \cdot |u_i|$. This implies a bound on $A_{i, D(u_i)}$ of at most $1 + 4 t_{D(u_i)} \cdot |u_i|$, as desired.
\end{proof}

Next we observe that the probability of any node $u_i$ with $i \ge \log \epsilon^{-1} + \omega(\log n)$ ever dying is very small.
\begin{lemma}
The probability of any node $u_i$ with $|u_i| \ge \Omega(\epsilon^{-1} \log^2 n)$ ever dying is at most $n^{-\omega(1)}$.
\label{lem:bignodesdontdie}
\end{lemma}
\begin{proof}
It suffices to show that each such node $u_i$ has probability at most $n^{-\omega(1)}$ of dying. In order for $u_i$ to die, there would need to be some phase $j$ in which one of $u_i$'s children is corrupted. In any phase $j$, the probability of this happening is at most
$$e^{-\Omega(t_j \cdot |u_i|)},$$
since, by a Chernoff bound for sampling without replacement, the probability of a child $x$ of $u_i$ having a number of free slots outside the interval $[0.9 t_j |x| + 1, 1.1 t_j |x| - 1]$ at the start of phase $j$ is $e^{-\Omega(t_j |x|)} = e^{-\Omega(t_j |u_i|)}$. Since $t_j \ge \epsilon$, this probability evaluates to $n^{-\omega(1)}$, as desired.
\end{proof}

Thus, to complete the proof of Proposition \ref{prop:geo}, it suffices to show that the random variable $$\sum_{i  \le \log (\epsilon^{-1} \log^2 n)} t_{D(u_i)} \cdot |u_i|$$ is at most $O(\log \epsilon^{-1})$ plus a geometric random variable with mean $O(1)$. We will prove this at the end of the section with Lemma \ref{lem:sumdeath}. Before we do so, however, we will need a somewhat subtle probability argument that begins with a sequence of simple claims about how different events relate to one another.

\begin{claim}
    Let $j < k \le o(\log n)$. Consider some node $u_i$, and let $S$ be a set of nodes that do not intersect with $u_i$ and that satisfy $\sum_{x \in S} |x| \le \sqrt{n}$. Define $E_1$ to be an arbitrary outcome for which slots are activated in the nodes in $S$ during phases $1, 2, \ldots, k$; and let $E_2$ be an outcome for which slots are activated in $u_i$ during phases $1, 2, \ldots, j$, subject to the condition that there are at most $2 |u_i| t_{j + 1}$ free slots in $u_i$ at the end of phase $j$. Finally, define $E_3$ to be the event that $u_i$ contains at least $4 |u_i| t_{k + 1}$ free slots at the end of phase $k$. Then,
    $$\Pr[E_3 \mid E_1, E_2] \le e^{-\Omega(t_{k + 1} \cdot |u_i|)}.$$
    \label{clm:helpercorruptiona}
\end{claim}
\begin{proof}
Let $q$ be the number of free slots in nodes outside of $S$ that remain unactivated at the end of phase $k$, conditioned on $E_1$ (which affects $q$ by at most $\pm O(\sqrt{n})$) and $E_2$ (which has no effect on $q$). Then,
$$q = t_{k + 1} n \pm O(\sqrt{n}) = (1 \pm o(1)) t_{k + 1} n,$$
where the final equality uses the fact that $k = o(\log n)$ and thus that $n/t_{k + 1} = n^{1 - o(1)}$.

Condition on $E_1, E_2$. Let $F$ be the free slots in $[n] \setminus \bigcup_{x \in S} x$ at the end of phase $j$; let $F' \subseteq F$ be the free slots in $u_i$ at the end of phase $j$; and let $Q$ be the $q$ free slots in $F$ that remain free at the end of phase $k$. Then $Q$ is a random subset of $F$ of size $q = (1 \pm o(1)) t_{k + 1} n$. Since $|F| = t_{j + 1} n$, it follows that $Q$ is a random subset of $F$ of size at at most
\begin{equation}
    (1 + o(1)) \frac{t_{k + 1}}{t_{j + 1}} |F|.
\label{eq:q-bound}
\end{equation}

By a Chernoff bound for sampling without replacement, if we define $\mu = \E[|Q \cap F'| \mid E_1, E_2]$, then we have for any $\Delta \ge \Omega(\mu)$ that
$$
\Pr[|Q \cap F'| \ge \mu + \Delta \mid E_1, E_2] \le e^{-\Omega(\Delta)}.
$$
Since $F' \le 2|u_i|t_{j + 1}$, we know from \eqref{eq:q-bound} that $\mu \le 2|u_j|t_{j + 1} \cdot (1 + o(1))\frac{t_{k + 1}}{t_{j + 1}} = (2 + o(1))|u_j|t_{k + 1}$. Thus we can set $\Delta = |u_j| t_{k + 1}$ to obtain
$$
\Pr[|Q \cap F'| \ge 4 |u_j| t_{k + 1} \mid E_1, E_2] \le e^{-\Omega(|u_j|t_{k + 1})}.
$$
This implies that
$$\Pr[E_3 \mid E_1, E_2] \le e^{-\Omega(t_{k + 1} \cdot |u_i|)},$$
as desired.
\end{proof}

Let $\gamma$ be a sufficiently small positive constant, and say that an interval $I \subseteq [n]$ is \defn{$\gamma$-imbalanced} at the end of phase $p$ if the number of free slots in $I$ is outside the interval $[t_j \cdot |I| \cdot (1 - \gamma) + 1, t_j \cdot |I| \cdot (1 + \gamma) - 1]$.

\begin{claim}
    Consider a phase $k = o(\log n)$. Consider some interval $I \subseteq [n]$. Let $S$ be a set of nodes that do not intersect with $I$ and that satisfy $\sum_{x \in S} |x| \le \sqrt{n}$. Define $E_1$ to be an arbitrary outcome for which slots are activated in the nodes in $S$ during phases $1, 2, \ldots, k$.
    Let $E_2$ be the event that, at the end of phase $k$, $I$ is $\gamma$-imbalanced. Then,
    $$\Pr[E_2 \mid E_1] \le e^{-\Omega(t_{k + 1} \cdot |I|)}.$$
    \label{clm:helpercorruptionb}
\end{claim}
\begin{proof}
Conditioned on $E_1$, let $q$ be the number of free slots in $[n] \setminus \bigcup_{x \in S} x$ that remain at the end of phase $k$. Since $\sum_{x \in S} |x| \le \sqrt{n}$, we have that $q = t_{k + 1} \cdot n \pm O(\sqrt{n}) = (1 \pm o(1)) t_{k + 1} n$.

Let $F = [n] \setminus \bigcup_{x \in S} x$, and let $Q \subseteq F$ be the $q$ slots in $F$ that remain free at the end of phase $k$. Then $Q$ is a random subset of $F$ of size
\begin{equation}(1 \pm o(1)) t_{k + 1} n = (1 \pm o(1)) t_{k + 1} |F|.
    \label{eq:q-bound-b}
\end{equation}

By a Chernoff bound for sampling without replacement, if we define $\mu = \E[|Q \cap I| \mid E_1]$, then we have for any $\Delta \ge \Omega(\mu)$ that
$$\Pr[||Q \cap I| - \mu| \ge \Delta \mid E_1] \le e^{-\Omega(\Delta)}.$$
By \eqref{eq:q-bound-b}, we have that $\mu = (1 \pm o(1)) t_{k + 1} |I|$, which supposing (wlog) that $n$ is large enough gives $|\mu - t_{k + 1} |I|| \le \gamma |I| t_{k + 1} / 2$. Thus we can set $\Delta = \gamma |I| t_{k + 1} / 2 - 1$ to obtain
$$\Pr[||Q \cap I| - t_{k + 1} |I| | \ge \gamma |I| t_{k + 1} / 2 - 1 \mid E_1] \le e^{-\Omega(|I|t_{k + 1})}.$$
This implies that
$$\Pr[E_2 \mid E_1] \le e^{-\Omega(t_{k + 1} \cdot |I|)},$$
as desired.
\end{proof}

\begin{claim}
    Consider some node $u_i$ and consider some $r \le i - 100$. Let $S$ be a set of nodes that do not intersect with $u_i$ and that satisfy $\sum_{x \in S} |x| \le \sqrt{n}$.

    Let $j < k \le o(\log n)$ be phase numbers. Condition on:
    \begin{enumerate}
    \item An arbitrary outcome $E_1$ for which slots are activated in the nodes in $S$ during phases $1, 2, \ldots, k$.
    \item Some history $E_2$ for which slots get activated in $u_r$ during phases $1, 2, \ldots, j$, subject to the condition that $u_r$ remained uncorrupted the end of phase $j - 1$.
    \end{enumerate}
    Finally, define $E_3$ to be the event that at least one of $u_i$'s children becomes corrupted at the beginning of phase $k + 1$. Then,
    $$\Pr[E_3 \mid E_1, E_2] \le e^{-\Omega(t_{k + 1} \cdot |u_i|)}.$$
    \label{clm:helpercorruptionc}
\end{claim}
\begin{proof}
Let $a$ be the child of $u_i$ not containing $u_r$ and let $b$ be the child containing $u_r$. Let $A$ denote the event that $a$ becomes corrupted at the beginning of phase $k + 1$ and let $B$ denote the event that $b$ becomes corrupted at the beginning of phase $k + 1$. By Claim \ref{clm:helpercorruptionb}, we have that
$$\Pr[A \mid E_1] \le e^{-\Omega(t_{k + 1} \cdot |a|)} = e^{-\Omega(t_{k + 1} \cdot |u_i|)}.$$

The only way for $b$ to become corrupted in phase $k$ is if at least one of the following two events occurs:
\begin{itemize}
\item Event $B_1$: $b \setminus u_r$ is $\gamma$-imbalanced by the end of phase $k$.
\item Event $B_2$: At the end of phase $k$, there are at least $4 t_{k + 1} |u_r|$ free slots in $u_r$.
\end{itemize}
 By Claims \ref{clm:helpercorruptiona} and \ref{clm:helpercorruptionb}, along with the fact that $|u_r|, |b| = \Theta(|u_i|)$, we have that
$$\Pr[E_3 \mid E_1, E_2] \le \Pr[B_1 \cup B_2 \mid E_1, E_4] \le e^{-\Omega(t_{k} \cdot |u_r|)}.$$
\end{proof}

\begin{lemma}
    Consider some node $u_i$, with $i \ge 2$, and consider some $r \le i - 100$. Let $S_1$ be a set of nodes that do not intersect with $u_i$ and that satisfy $\sum_{x \in S_1} |x| \le \sqrt{n}$. Let $S_2$ be a set of nodes contained in $u_i \setminus u_r$ that satisfy $|x| \ge \Omega(|u_i|)$ for each $x \in S_2$.

    Let $j < k$ be phase numbers. Condition on:
    \begin{enumerate}
    \item No node in $S_1$ being corrupt at the start of phase $k$. (Call this event $E_1$.)
    \item  $u_r$'s children remaining uncorrupted at the start of phase $k$. (Call this event $E_2$.)
    \item No node in $S_2$ being corrupt at the start of phase $j + 1$. (Call this event $E_3$.)
    \item  Some history $E_4$ for which slots get activated in $u_r$ during phases $1, 2, \ldots, j$, subject to the condition that $u_r$ remained uncorrupted at the start of phase $j + 1$.
    \end{enumerate}
    Then, conditioned on $E_1, E_2, E_3, E_4$, the conditional probability of either of $u_i$'s children becoming corrupted  at the beginning of phase $k + 1$ is at most
    $$e^{-\Omega(t_{k + 1} \cdot |u_i|)}.$$
    \label{lem:conditionalcorruption1}
\end{lemma}
\begin{proof}
Let $E_5$ denote the event that one of $u_i$'s children becomes corrupted at the beginning of phase $k + 1$. By Claim \ref{clm:helpercorruptionc}, we have that
$$\Pr[E_5 \mid E_1, E_4] \le e^{-\Omega(t_{k + 1} \cdot |u_i|)}.$$

We also have by Claim \ref{clm:helpercorruptionc} that
$$\Pr[E_2 \mid E_1, E_4] \le \sum_{q = 1}^k e^{-\Omega(t_q \cdot |u_r|)} \le e^{-\Omega(t_k \cdot |u_r|)},$$
and by Claim \ref{clm:helpercorruptionb} that
$$\Pr[E_3 \mid E_1, E_4] \le \sum_{x \in S_2} \sum_{q = 1}^{j + 1} e^{-\Omega(t_q \cdot |x|)} \le O(1) \cdot \sum_{q = 1}^k e^{-\Omega(t_q \cdot |u_r|)} \le e^{-\Omega(t_k \cdot |u_r|)}.$$

Therefore,
\begin{align*}
& \Pr[E_5 \mid E_1, E_2, E_3, E_4] \\
& \le \Pr[E_5 \mid E_1, E_4] + \Pr[E_2 \mid E_1, E_4] + \Pr[E_3 \mid E_1, E_4] \\
& \le e^{-\Omega(t_{k + 1} \cdot |u_i|)},
\end{align*}
as desired.
\end{proof}

Finally, our main lemma will make use of the following simple fact:
\begin{claim}
If a non-leaf $u$ becomes corrupt at the start of phase $j$, then at least one of its children must also be corrupt by the start of phase $j$.
\label{clm:childcorruption}
\end{claim}
\begin{proof}
For $u$ to become corrupt its number of free slots must exit the interval $[t_j \cdot |u| \cdot 0.9 + 1, t_j \cdot |u| \cdot 1.1 - 1]$. For this to happen, at least one of $u$'s children $a$ must have a number of free slots outside the interval $[t_j \cdot |a| \cdot 0.9 + 1/2, t_j \cdot |a| \cdot 1.1 - 1/2]$. Thus $u$ has at least one child that is corrupt by the end of phase $j$.
\end{proof}

We are now prepared to prove the main lemma of the section. The following lemma, when combined with Lemmas \ref{lem:bignodesdontdie} and \ref{lem:remainingslotsbound}, implies Proposition \ref{prop:geo}.
\begin{lemma}
Suppose that $\epsilon^{-1} = n^{o(1)}$, and set $\tau = \log (\epsilon^{-1} \log^2 n)$. Then, the random variable $$\sum_{i  \le \tau} |u_i| \cdot t_{D(u_i)}$$
is at most $O(\log \epsilon^{-1})$ plus a geometric random variable with mean $O(1)$.
\label{lem:sumdeath}
\end{lemma}
\begin{proof}
Let $q_i$ be the largest $q \le \log (\epsilon^{-1})$ such that $u_q$ is dead by the start of phase $i + 1$. Define $z_i$ (with $z_0 = 0$) recursively as follows: For $i > 0$, if $z_{i - 1} \ge q_i$, then $z_i = z_{i - 1}$; otherwise, $z_i = q_i + 100$.

Also define $d_i$ (with $d_0 = 0$) recursively as follows: For $i > 0$, if $z_i = z_{i - 1}$, then $d_i = 0$; otherwise, $d_i = |u_{z_i}| \cdot t_{i + 1}$.

Observe that
$$\sum_{i \ge 0} |u_i| \cdot t_{D(u_i)} \le \sum_{i \ge 0} d_i,$$
so it suffices to show that $\sum_{i \ge 0} d_i$ is at most a geometric random variable with mean $O(\log \epsilon^{-1})$.

We will prove that, if we condition on any set of outcomes for $z_1, z_2, \ldots, z_{i - 1}$, then the conditional distribution of $d_i$ is stochastically dominated by a geometric random variable with mean $O(1)$. It follows that $\sum_i d_i$ is stochastically dominated by a sum of $O(\log \epsilon^{-1})$ independent geometric random variable each of mean $O(1)$, which implies the lemma by a Chernoff bound for sums of independent geometric random variables.

For the rest of the proof, condition on any set of outcomes for $z_1, z_2, \ldots, z_{i - 1}$. Let $z:= z_{i - 1}$, and let $j < i$ be the smallest $j$ such that $z_j = z_{i - 1}$. Fix some $s \in [z + 1, \log \epsilon^{-1} + \log^2 n]$. We will seek to bound the probability of $u_s$ dying in phase $i + 1$ to be at most $e^{-\Omega(t_{i + 1}\cdot |u_s|)}$. Note that, if $|u_s| \in O(1)$, then this bound is trivial, so we may assume without loss of generality that $s > 100$.

As helper notation, for $r \ge 1$, let $x_r$ denote the child of $u_r$ that does not contain $u_0$ (i.e., the child that is not $u_{r - 1}$).

In order to use Lemma \ref{lem:conditionalcorruption1}, it is helpful to note that conditioning on $z_1, z_2, \ldots, z_{i - 1}$ is equivalent to conditioning on:
\begin{enumerate}
\item The event $E_1$ that none of $u_s, u_{s + 1}, \ldots, u_{\tau}$ die by the beginning of phase $i$. By Claim \ref{clm:childcorruption}, this is equivalent to the event that: none of $x_{s + 1}, \ldots, x_{\tau}$ corrupt by the beginning of phase $i$, and neither of $u_s$'s children corrupt by the beginning of phase $i$.
\item The event $E_2$ that, during phases $1, 2, \ldots, j$ (including the beginning of phase $j + 1$), none of the nodes $u_{s - 99}, u_{s - 98}, \ldots, u_{s}$ die. This is equivalent to saying that none of the children of the nodes $u_{s - 99}, u_{s - 98}, \ldots, u_s$ become corrupt by the beginning of phase $j + 1$; which, by Claim \ref{clm:childcorruption}, is equivalent to saying that none of the nodes $x_{s - 99}, x_{s - 98}, \ldots, x_s$ or $u_{s - 100}$ become corrupt by the end of phase $j + 1$.
\item The event $E_3$ that, during phases $k \in [j]$, the free-slot activations in $u_0, \ldots, u_{s - 100}$ are consistent with the sequence $z_1, z_2, \ldots, z_j$, including the fact that node $u_{s - 100}$ remains uncorrupted at the beginning of phase $j + 1$.
\end{enumerate}

In order for $u_s$ to die in phase $i$, one of its children must become corrupt. Notice, however, that the above conditions meet the criteria for Lemma \ref{lem:conditionalcorruption1} (with $u_s$ serving the role of $u_i$ in the lemma).\footnote{Note that Lemma \ref{lem:conditionalcorruption1} requires the number of phases $k$ to be $o(\log n)$, which holds here because $\epsilon^{-1} = n^{o(1)}$. It also requires that $\sum_{x \in S_2} |x| \le \sqrt{n}$, which holds here because $S_2$ is a set of at most $O(\log \epsilon^{-1})$ nodes each of which are size at most $O(2^\tau) = O(\epsilon^{-1} \log^2 n) = n^{o(1)}$.}  We can therefore deduce that the conditional probability of $u_s$ dying at the start of phase $i + 1$ is at most
$$e^{-\Omega(t_{i + 1} \cdot |u_s|)}.$$

Thus the distribution of $d_i$ satisfies
$$\Pr[d_i \ge t_{i + 1} \cdot |u_s|] \le \sum_{j \ge s} e^{-\Omega(t_{i + 1} \cdot |u_j|)} \le e^{-\Omega(t_{i + 1} \cdot |u_s|)},$$
which implies that the conditional distribution of $d_i$ is stochastically dominated by a geometric random variable with mean $O(1)$, as desired.
\end{proof}

\subsection{A Matching Lower Bound for Insertion Algorithms Without Evictions}\label{sec:matchinglowerbound}

In this section, we show that the query time of $O(\log \epsilon^{-1})$ achieved by Theorem \ref{thm:non-greedy} is tight for any two-choice insertion algorithm that does not perform evictions.

\begin{theorem}
    Any two-choice insertion algorithm that does not perform evictions must incur an expected query time of $\Omega(\log \epsilon^{-1})$ at load factor $1 - \epsilon$.
\end{theorem}
\begin{proof}
Consider the time to query a random element $u$ out of those present in the hash table. As noted in \cite{Peterson57}, this quantity is agnostic to how we order elements within the hash table, so long as we keep each element $u$ at or to the right of its hash $h(u)$, and so long as we do not place any free slots between $h(u)$ and the position containing $u$. Thus, we may assume without loss of generality that the hash table is unordered, and that each insertion simply places the element $u$ in the first free slot to the right of its hash $h(u)\in \{h_1(u), h_2(u)\}$.

Now, consider an element $u$ that was inserted at load factor $1 - \delta$ for some $\delta$. Consider the state of the hash table prior to $u$'s insertion, and let $F$ be the number of free slots in the intervals $[h_1(u), h_1(u) + \delta^{-1}/4)$ and $[h_2(u), h_2(u) + \delta^{-1}/4)$. Since the load factor is $1 - \delta$, we have that $\E[F] \le 1/2$, which implies by Markov's inequality that $\Pr[F \ge 1] \le 1/2$ and thus that $\Pr[F = 0] \ge 1/2$. If $F = 0$, then $u$ is placed distance at least $\delta^{-1}/4$ from its hash $h(u)$, and thus has query time $\Omega(\delta^{-1})$. Since this happens with probability $\Omega(1)$, the expected query time for $u$ is $\Omega(\delta^{-1})$.

Finally, averaging over all $u$, the expected time to query a random element out of those present is
$$\int_{\delta = \epsilon}^1 \Omega(\delta^{-1}) \, d \delta = \Omega(\log \epsilon^{-1}),$$
as desired.
\end{proof}

\section{A Two-Choice Algorithm With Evictions, and with $O(1)$ Queries}\label{sec:O1queries}

Finally, in this section, we turn out attention to insertion algorithms that are allowed to change which hash function $h \in \{h_1, h_2\}$ elements $y$ that are already in the hash table are using. This is to say that the insertion algorithm can perform \emph{evictions}, moving an element $y$ between the region containing $h_1(y)$ and the region containing $h_2(y)$. We prove that such insertion algorithms can achieve a remarkably strong guarantee: an expected insertion time of $O(\epsilon^{-1/2})$ and an expected query time of $O(1)$.

\thmEvictions*

We remark that, as in Section \ref{sec:nongreedy}, we will assume that we can keep track of $O(1)$ bits of metadata per slot. As before, this assumption is without loss of generality in the sense that one can straightforwardly encode such information implicitly by slightly reordering elements.

\subsection{The Algorithm}

The algorithm proceeds in two phases.

\paragraph{Phase 1: Fill to load factor $1 - \epsilon_0$. }Let $\epsilon_0 \in (0, 1)$ be a sufficiently small positive constant (not an asymptotic parameter). For the first $(1 - \epsilon_0)n$ insertions, the algorithm uses standard ordered linear probing, placing each element $x$ using its first hash $h_1(x)$. As we will see, the point of this phase is to ``warmup the system'', so that there are many elements already present using their first hash.

\paragraph{Phase 2: Dart throwing. } After Phase 1, the remaining insertions use a more interesting insertion algorithm. As in Section \ref{sec:nongreedy}, we will assume in the algorithm description that we can keep track of $O(1)$ bits of metadata information per slot -- this assumption is without loss of generality in the sense that one can straightforwardly encode such information implicitly by slightly reordering elements.

Suppose, at some given moment that $\epsilon n$ free slots $f_1 < f_2 < \cdots < f_{\epsilon n}$ remain. Then, we will keep track of what we call the \defn{target region} of the hash table, which can be defined as follows. For each free slot $f_i \in \{f_1, f_2, \ldots, f_{\epsilon n}\}$ (where we consider $f_i$ after considering $f_1, f_2, \ldots, f_{i - 1}$) define the \defn{target slots} $s_1(i), s_2(i), \ldots, s_{\lceil \sqrt{\epsilon^{-1}} \rceil}(i)$ to be the slots in the hash table that minimize $f_i - s_j \pmod n$ (i.e., they are as close on $f_i$'s left as possible, with wraparound) out of the slots that are not already target slots for $f_1, f_2, \ldots, f_{i - 1}$. For now, we will assume that we already know which slots are target slots, and then later on we will discuss how to maintain this information.

To perform an insertion of an element $x$ during Phase 2, we use the following procedure to insert $x$ using its second hash $h_2(x)$:
\begin{itemize}
    \item \textbf{Case 1. If $h_2(x)$ is a target slot for some $f_i$: } In this case, we will perform the insertion in a way that fills in slot $f_i$, does not change the status of any other free slot, and keeps the elements in the hash table in sorted order by hash.

    Specifically, we can do this as follows. Identify the first position $j \ge h_2(x)$ that contains an element $y$ satisfying $h(y) > h_2(x)$ (where $h$ denotes whichever hash function $y$ is currently using). If $j < f_i$, then let $y_1, y_2, \ldots, y_t$ denote the elements between $y$ and slot $f_i$ (from left to right and inclusive), place $y_t$ in position $f_t$, place each $y_i$ with $i < t$ in the position previously occupied by $y_{i + 1}$, and place $x$ in the position previously occupied by $y_1$. Conversely, if $j > f_i$, then let $y_1, y_2, \ldots, y_t$ denote the elements after slot $f_i$ and before $y$ (from left to right), move element $y_1$ to slot $f_i$, move each $y_i$ with $i > 1$ to the position previously occupied by $y_{i - 1}$, and place $x$ in the position previously occupied by $y_t$ (if no $y_i$s exist, just place $x$ in $f_i$).\footnote{Note that, in this case where the algorithm moves elements to the left, it only moves elements with hashes $\le h_2(x)$ to positions $\ge h_2(x)$, which keeps each element in a legal position.}

    \item \textbf{Case 2. If $h_2(x)$ is not a target slot: } In this case, we will find an element $z$ close to $h_2(x)$ that is currently using its \emph{first hash}, we will evict $z$ to use its second hash (inserting $z$ recursively just as we are inserting $x$ now), and we will then insert $x$ in such a way that $z$'s freed slot gets filled, no other free slots get filled, and the elements in the hash table remain ordered by hash.

    Specifically, we accomplish all of this as follows. Identify the first position $j \ge h_2(x)$ that contains an element $y$ satisfying $h(y) \ge h_2(x)$ (where $h$ denotes whichever hash function $y$ is currently using). Let $z$ be the first element to appear at or after position $j$ such that $z$ is using its first hash. Let $y_1, y_2, \ldots, y_t$ denote the elements between $y$ and $z$ inclusive, listed left to right with $y_1 = y$ and $y_t = z$ (note that $y$ may equal $z$). Then, \defn{evict} $y_t = z$, and for each $i \in t - 1, t -2 , \ldots, 1$, place $y_i$ in the position that $y_{i + 1}$ formerly occupied. Place $x$ in the position $j$ formerly occupied by $z$.  Finally, take the evicted element $z$ and attempt to reinsert it using its second hash (i.e., $z$ now takes the role of $x$).
\end{itemize}

As discussed above, the insertion process is recursive. Whenever we are in Case 2 ($h_2(x)$ is not a target slot), we find some nearby element $z$ that is using its first hash, and we \emph{evict} $z$, reinserting it from scratch using its second hash. Each time during this recursive process that we attempt to insert some element $z$ using its second hash, we think of this as a \defn{dart throw}, where we are throwing a random dart (namely $h_2(x)$) at the hash table, in hopes of hitting a target hash.

The final dart throw is the one that finds a target slot, and uses some free slot $f_i$. After this dart throw, the final step in the insertion algorithm is to update which slots are target slots. This can be done in time $O(|T|)$, where $T$ is the (maximal-length) interval of target slots containing $f_i$. We will, as part of the analysis, show that $\E[|T|] = O(\sqrt{\epsilon^{-1}})$.

In addition to keeping track of the target slots \emph{during} Phase 2, we must compute \emph{at the end of Phase 1} who the initial target slots are. This can be done in time $O(n)$, contributing $O(1)$ amortized time to each operation in Phase 1. This time bound can then be deamortized in any of several standard ways -- for example, by randomizing $\epsilon_0$ in some range of size $\Theta(1)$, we can randomize when the $O(n)$ work occurs so that each operation takes worst-case expected time $O(1)$.

\paragraph{A remark on failures. }One feature of the above algorithm is that, if \emph{every} Phase 1 element is moved to use its second hash then the algorithm simply fails -- the insertion protocol in Case 2 never finds any $z$ that it can evict. We will show that this happens with a negligible probability of $1 / \poly(n)$. Since the failure mode is so rare, we can feel free to do essentially anything we want to resolve it (e.g., allow all remaining operations to take $n$ time). Once we prove that, with high probability, the failure event does not occur, we will then ignore the failure case in our analysis.

\paragraph{Analysis overview. } Before continuing, let us take a moment to explain, at least intuitively, why the above algorithm ends up achieving an expected query time of $O(1)$.

For an insertion at load factor $1- \epsilon$, the fraction of slots that are target slots is $\sqrt{\epsilon}$, so the expected number of dart throws needed to find a target slot is $\sqrt{\epsilon}^{-1}$. Summing over the insertions in Phase 2, the expected \emph{total} number of dart throws is
$$n \cdot \int_0^{\epsilon_0} \sqrt{\epsilon} \, d \epsilon =  \frac{2n}{3} \epsilon_0^{3/2}.$$
Selecting $\epsilon_0$ to be a small enough constant, this is at most, say, $n / 100$.

Why does the number of dart throws matter? It controls what fraction of elements use their \emph{first hash}. If there are only $n/100$ total dart throws, then \emph{most elements} still use their first hash at the end of the construction. This means that, intuitively, each call to Case 2 above should be able to complete in $O(1)$ expected time.

In total, we perform Case 2 at most $n/100$ times (at most once for each dart throw). If each instance of Case 2 takes $O(1)$ expected time, then it should increase $Q$ by $O(1)$ in expectation. The total expected increase $Q$ from Case 2, over all insertions, is therefore $O(n)$.

Note that, here, something very interesting has already happened. Each dart throw has a $\sqrt{\epsilon}$ probability of success (much better than $\epsilon$!), but failed dart throws \emph{do not} cost $\sqrt{\epsilon}^{-1}$, they cost $O(1)$! This is where we save a roughly $\sqrt{\epsilon}^{-1}$ factor when compared to standard insertion algorithms.

What about the increase in query costs from Case 1? Here, there is a failure mode we must be careful to avoid: that free slots might ``cluster together'' (as in standard linear probing) to create large regions of target slots in which Case 1 is expensive. Observe, however, that our algorithm avoids this by design -- it maintains the invariant that the current set of free slots, at any moment in Phase 2, is a uniformly random subset of those present at the end of Phase 1. This allows us to argue that target regions (regions of consecutive target slots) tend to contain just $O(1)$ free slots, and that when Case 1 occurs, it tends to increase $Q$ by only $O(\sqrt{\epsilon}^{-1})$. Since Case 1 occurs only once per insertion, we can bound the total expected increase to $Q$ from insertions performing Case 1 by
$$n \cdot \int_0^{\epsilon_0} O(\sqrt{\epsilon}^{-1}) \, d \epsilon =  O(n).$$

We will see in the next subsection how to formalize these arguments, showing that $\E[Q] = O(n)$ after all the insertions are complete. In fact, we will prove stronger claims, showing that an insertion at load factor $1 - \epsilon$ takes $O(\sqrt{\epsilon^{-1}})$ expected time (faster than the standard bound of $\epsilon^{-1}$!), and that the expected time to perform \emph{any} query (positive or negative) is $O(1)$.

\subsection{Analysis}

We begin by making some basic observations about the algorithm which we will then employ implicitly throughout the analysis:
\begin{enumerate}
    \item At any given moment, all elements using their first hash were inserted in Phase 1.
    \item Elements that are using their first hash can be evicted to use their second, but not vice-versa. Moreover, when an element using its first hash is evicted, its second hash is uniformly random and independent of all other hashes used to construct the hash table so far.
    \item  Because each free slot has the same number of target slots assigned to it, each insertion in Phase 2 uses a uniformly random free slot.
\end{enumerate}

We now begin the analysis. Define a \defn{target zone} to be a maximal sequence of adjacent target slots (possibly with wraparound). We begin by arguing that, for a given element $x$, the expected size of the target zone containing $h_2(x)$ is $O(1)$. Note that, if there are $\epsilon n$ free slots, then only an $O(\sqrt{\epsilon})$ fraction of slots overall will be in target zones, so an immediate consequence of Lemma \ref{lem:targetzone} is that, for slots which \emph{are} in target zones, the average target zone size is $O(\sqrt{\epsilon}^{-1})$.
\begin{lemma}
Let $x$ be an element (possibly in the table), and suppose the load factor is $1 - \epsilon$, with $\epsilon \le \epsilon_0$. Let $h \in \{h_1(x), h_2(x)\}$. Let $T$ be the target zone containing $h$ (or the empty interval if $h$ is not a target slot). Then, $\E[|T|] \le O(1)$.
\label{lem:targetzone}
\end{lemma}
\begin{proof}
For $h$ to be in a target zone of size $j$, there must be an interval of the form $I = [h -a, h + b - 1]$, with $a + b = j$, such that $I$ contains at least $\lceil \sqrt{\epsilon}j \rceil$ free slots. Rounding $a$ and $b$ up to multiples of $\sqrt{\epsilon}^{-1}$, we see that there must exist $a', b'$ which are multiples of $\sqrt{\epsilon}^{-1}$ such that $a' + b' \in [j, j + 2\sqrt{\epsilon}^{-1}]$ and $[h -a', h + b' - 1]$ contains at least $\lceil \sqrt{\epsilon}j \rceil =  (a' + b')\sqrt{\epsilon} / 3$ free slots. Therefore, the size of the target zone is at most the maximum value of $b' + a'$ over all non-negative multiples $a',b'$ of $\sqrt{\epsilon^{-1}}$ such that $[h -a', h + b' - 1]$ contains at least $(a' + b') \sqrt{\epsilon} / 3$ free slots.

Fix some $a'$ and $b'$ that are multiples of $\sqrt{\epsilon}^{-1}$ and such that $a' + b' \in [j, j + 2 \sqrt{\epsilon}^{-1}]$. Let $F$ be the $\le a' + b'$ free slots in $[h_2(x) - a', h_2(x) + b' - 1]$ at the end of Phase 1, and condition on $F$ (it is no longer random). Each insertion in Phase 2 (except possibly the one that either inserts $x$ or evicts $x$ to its second hash) uses a uniformly random free slot. So, out of all $\epsilon_0 n + 1$ free slots that existed at the end of Phase, at most a uniformly random subset of size $\epsilon n + 1$ now remain (the $+1$ is to allow for the fact that $x$ may have used a non-random free slot). If we define $F'$ to be the set of free slots in $F$ that remain, then $|F'|$ is at most a sum of $|F| \le a' + b'$ negatively associated indicator random variables, each of which is $1$ with probability at most $(\epsilon + 1/n) / \epsilon_0 = O(\epsilon)$. By a Chernoff bound, $\Pr[|F'| \ge J \cdot \E[F']] \le J^{-\Omega(J \cdot \E[F'])}$ (see, e.g., \cite{DBLP:conf/sosa/Kuszmaul25}), so
$$\Pr[|F'| \ge (a' + b')\sqrt{\epsilon}/3] \le \epsilon^{-\Omega((a' + b') \sqrt{\epsilon})}.$$
Since the expected value of $|F'|$ is at most $\epsilon (a' + b')$, we also have by Markov's inequality that $\Pr[|F'| \ge 1] \le \epsilon \cdot (a' + b')$. So,
$$\Pr[|F'| \ge (a' + b')\sqrt{\epsilon} / 3] \le \min(\epsilon \cdot (a' + b'), \epsilon^{-\Omega((a' + b') \sqrt{\epsilon})}).$$

Recall that the size of the target zone is at most the maximum value of $b' + a'$ over all non-negative multiples $a',b'$ of $\sqrt{\epsilon^{-1}}$ such that $[h -a', h + b' - 1]$ contains at least $(a' + b') \sqrt{\epsilon} / 2$ free slots. For each $r \ge 1$, there are $O(r)$ values of $a', b'$ that are multiples of $\sqrt{\epsilon}^{-1}$ with $a' + b' = r \sqrt{\epsilon}^{-1}$. Thus, the expected size of the target zone is at most
$$\sum_{r \ge 1} O(r \cdot \sqrt{\epsilon}^{-1}) \cdot \min(\sqrt{\epsilon} r, \epsilon^{-\Omega(r)}) = O(1).$$
\end{proof}

Next we bound the total number of dart throws, over all time, with high probability. This lemma plays a critical role: it says that most of the elements inserted in Phase 1 \emph{never} get evicted. Intuitively, this is the reason that Case 2 of the insertion protocol can hope to be efficient (expected time $O(1)$).
\begin{lemma}
With high probability in $n$, the total number of dart throws is at most $n / 100$.
\label{lem:countdarts}
\end{lemma}
\begin{proof}
The insertions in Phase 1 do not involve dart throws. For $\epsilon \le \epsilon_0$, during the $(1 - \epsilon)n + 1$-th insertion, each dart throw succeeds with probability at least $\sqrt{\epsilon}$. The total number of dart throws is therefore stochastically dominated by
$$\sum_{j = 1}^{\epsilon_0 n} Y_{j},$$
where each $Y_{j}$ is independently a geometric random variable with mean $\sqrt{n/j}$. The expected value is at most
\begin{align*}
\sum_{j=1}^{\epsilon_0 n} \left(\sqrt{j/n}\right)^{-1} \le n\int_0^{\epsilon_0}\epsilon^{-1/2}\,d\epsilon = 2n\sqrt{\epsilon_0} \le \frac{n}{200}.
\end{align*}

Each $Y_j$ is, with high probability in $n$, at most $O(\sqrt{n/j} \log n)$, so by Hoeffding's inequality,
$$\Pr\left[\sum_{j = 1}^{\epsilon_0 n} Y_j \ge n/200 + k\right] \le e^{-\Omega(k^2 / \sum_{j =   1}^{\epsilon_0 n} (\sqrt{n/j} \log n)^2)}.$$
Note that $\sum_{j = 1}^n (\sqrt{n/j} \log n)^2 = O(\log^2 n) \cdot \sum_{j = 1}^n O(n/j) \le O(n \log^3 n)$. So,
$$\Pr\left[\sum_{j = 1}^{\epsilon_0 n} Y_j \ge n/200 + k\right] \le e^{-\Omega(k^2 / (n \log^3 n))}.$$
Setting $k = n / 200$ completes the proof.
\end{proof}

For any position $j$, let $A(j)$ be the number of times in Phase 2 that a free slot $b > j$ (with wrap-around) has been filled by an element hashing (via its second hash) to a position $a \le j$ (with wrap-around). The quantity $A(j)$ captures one of two natural ways that an element $x$ with $h_2(x) = j$ can get moved to the right over time (namely, $A(j)$ captures moves due to Case 1 of the insertion protocol). And, critically for our analysis, $A(j)$ captures the \emph{only} way that an element using its first hash can move to the right over time (because such elements do not get moved around by Case 2 of the insertion protocol).

\begin{lemma}
    Let $x$ be an element (maybe in the hash table) and let $h \in \{h_1(x), h_2(x)\}$.
    Then $\E[A(h)] = O(1)$.
    \label{lem:Ah}
\end{lemma}
\begin{proof}
During the $(1 - \epsilon)n + 1$-th insertion, the expected size of the target zone containing $h$ is $O(1)$ (see Lemma \ref{lem:targetzone}), while the sum of the sizes of all target zones is $\Theta(\sqrt{\epsilon} n)$. Thus the probability of the target zone being used is $O(1 / (\sqrt{\epsilon} n))$. The probability of $A(h)$ being incremented by the insertion is therefore also $O(\sqrt{\epsilon^{-1}} / n)$. This implies
$$\E[A(h)] \le  O\left(\int_{\epsilon = 0}^1 \sqrt{\epsilon^{-1}} \, d\epsilon \right) = O(1).$$
\end{proof}

The most difficult step in the analysis will be to bound the quantity $j_2$ defined in the following lemma: The first position $j_2 \ge h_i(x)$, for some element $x$ and some hash function $h_i$, to contain an element $z$ whose first hash is greater than $h_i(x)$ and who is \emph{currently using} its first hash. Note that the $j_2 - h_i(x)$ serves as an upper bound for several quantities that we care about, including (1) the number of probes $h_i(x), h_i(x) + 1, \ldots$ that a query to $x$ would need to make, and (2) when $h_i = h_2$, the time needed to perform Case 2 of the insertion protocol on $x$.

To bound $j_2$, we begin in the following lemma with a (non-probabilistic) bound that relates $j_2$ to several other quantities.
\begin{lemma}
    Let $x$ be an element (maybe in the hash table) and let $h \in \{h_1(x), h_2(x)\}$. Let
    \begin{itemize}
    \item $j_1$ be the largest hash $j_1 \le h$ such that there is an element $y$ with hash $h_1(y) = j_1$ and such that $y$ is using its first hash.
    \item $j_2$ be the smallest position $j_2 \ge h$ to contain an element $z$ using its first hash, and where $h_1(z) > h$.
    \item $A_0$ be the number of elements that, at the end of Phase 1, sit in positions $> h$ but have (first) hashes $\le h$ (with wrap-around).
    \item $k_1(j_1, j_2)$ be the number of elements with first hashes in $(j_1, j_2)$ after Phase 1.
    \item $k_2(j_1, j_2)$ be the number of free slots in positions $[j_1, j_2]$ after Phase 1;
    \item $k_3(j_1, j_2)$ be the number of elements $z$ \emph{currently} using their second hash, and with $h_2(z) \in [j_1, j_2]$.
    \end{itemize}
    Then,
     $$j_2 \le h + A_0 + A(h) - k_1(j_1, j_2) + 2k_2(j_1, j_2) + 2k_3(j_1, j_2).$$
     \label{lem:k123}
\end{lemma}
\begin{proof}
Let $y$ be the right-most element with $h_1(y) = j_1$ and that is currently using its first hash, and let $z$ be the element in position $j_2$ (these are the same elements as in the definitions of $j_1$ and $j_2$). By construction, all elements that currently reside between $y$ and $z$ are using their second hash. At the end of Phase 1, there were at least $k_1$ elements using their first hashes that were between $y$ and $z$. These elements have all been evicted by the at-most $k_3$ elements that are now between $y$ and $z$ and using their second hashes. Thus $k_3 \ge k_1$. So, it suffices to show that
    $$j_2 \le   h + A_0 + A(h) + 2k_2(j_1, j_2) + k_3(j_1, j_2).$$
We also have $A(j_1) \le k_2(j_1, j_2) + A(h)$, since any insertion that contributes to $A(j_1)$ but not to $A(h)$ would need to use a free slot in $(j_1, h] \subseteq [j_1, j_2]$, so it suffices to show
$$j_2 \le h + A_0 + A(h) + k_2(j_1, j_2) + k_3(j_1, j_2).$$
At the start of Phase 2, $y$ is in position at most $h + A_0$. During Phase 2, $y$ is moved to the right at most $A(j_1)$ times (and may also be moved to the left by some insertions), so $y$'s current position is at most $h + A_0 + A(j_1)$. (Note that, since $y$ is using its first hash, it has not been moved by any invocations of Case 2 of the insertion protocol.) By construction, if $p(y)$ is the current position of $y$, then all slots in the interval $(p(y), j_2]$ are either free slots in the range $[j_1, j_2]$ (of which there are at most $k_2(j_1, j_2)$) or elements $z$ using their second hash and with $h_2(z) \in [j_1, j_2]$ (of which there are at most $k_3(j_1, j_2)$). Therefore, the number of such slots is at most $k_2(j_1, j_2) + k_3(j_1, j_2)$, which means that
$$j_2 \le h + A_0 + A(h) + k_2(j_1, j_2) + k_3(j_1, j_2),$$
as desired.
\end{proof}

To bound $j_2$ from the previous lemma, we first argue that for any fixed pair of values $j_1, j_2$, the quantity $2k_2(j_1, j_2) + 2k_3(j_1, j_2) - k_1(j_1, j_2)$ is very unlikely to be significantly positive.

\begin{lemma}
        Let $x$ be an element (maybe in the hash table) and let $h = h_2(x)$. Let $j_1\le h$ and let $j_2 \ge h$. Define $k_1(j_1, j_2)$, $k_2(j_1, j_2)$, and $k_3(j_1, j_2)$ as in Lemma \ref{lem:k123}. Then,
     $$\E[\max(0, 2k_2(j_1, j_2) + 2k_3(j_1, j_2) - k_1(j_1, j_2))] \le 1 / 2^{\Omega(j_2 - j_1)} + 1/ \poly(n).$$
     \label{lem:maxk123}
\end{lemma}
\begin{proof}
Observe that, for $r > 0$, we have:
\begin{enumerate}
    \item $k_2(j_1, j_2)$ satisfies $\Pr[k_2(j_1, j_2) \ge (j_2 - j_1)/10 + r] \le e^{-\Omega(r)}$. Recall that $k_2(j_1, j_2)$ is the number of free slots in the interval $[j_1, j_2]$ after Phase 1, at which point the hash table is a standard ordered linear-probing hash table at load factor $1 - \epsilon_0$. The number of such free slots is at most $|[j_1, j_2]| - X + Y$, where $X \sim \text{Bin}(n - 1, (1 - \epsilon_0)|[j_1, j_2]|/n)$ is the number of elements (besides $x$) with first hashes in $[j_1, j_2]$, and $Y$ is the number of elements with first hashes in $[j_1, j_2]$ that reside outside of positions in $[j_1, j_2]$ at the end of Phase 1. By the standard analysis of linear probing (see, e.g., Corollary 2 of \cite{bender2022linear}), $Y$ is stochastically dominated by a geometric random variable with mean $O(\epsilon_0^{-1}) = O(1)$. By a Chernoff bound on $X$, we have that $\Pr[X \le 0.9 \cdot |[j_1, j_2]| - r] \le e^{-\Omega(r)}$. Thus, $\Pr[|[j_1, j_2]| - X + Y + 1 \ge (j_2 - j_1)/10 + r] \le e^{-\Omega(r)}$.
    \item $k_3(j_1, j_2)$ satisfies $\Pr[k_3(j_1, j_2) \ge (j_2 - j_1)/10 + r] \le e^{-\Omega(r)} + 1/ \poly(n)$. Recall that $k_3(j_1, j_2)$ is the number of elements $z$ currently using their second hash, and with $h_2(z) \in [j_1, j_2]$. Also recall that, with high probability in $n$, there are at most $n/100$ total dart throws (Lemma \ref{lem:countdarts}). Besides the throw involving $x$'s second hash, each throw moves an element $y$ to a uniformly random second hash. The number of such $y$s whose second hashes are in $[j_1, j_2]$ is therefore, with high probability in $n$, bounded above by a sum of $n/100$ independent 0-1 random variables with total mean $(j_2 - j_1 + 1) / 100$. The probability of this sum evaluating to at least $(j_2 - j_1) / 10 + r - 1$ is, by a Chernoff bound, at most $e^{-\Omega(r)}$.
    \item $k_1(j_1, j_2)$ satisfies $\Pr[k_1(j_1, j_2) \le 0.9 (j_2 - j_1) - r] \le e^{-\Omega(r)}$. Recall that $k_1(j_1, j_2)$ is the number of elements with first hashes in $(j_1, j_2)$ after Phase 1. The expected number of such elements (besides $x$) is $(1 - \epsilon_0) (j_2 - j_1 - 1) \ge 0.95 (j_2 - j_1) - 1$. Thus, by a Chernoff bound, we have that $\Pr[k_1(j_1, j_2) \le 0.9 (j_2 - j_1) - r] \le e^{-\Omega(r)}$.
\end{enumerate}
So, $X = 2k_2(j_1, j_2) + 2k_3(j_1, j_2) - k_1(j_1, j_2)$ satisfies
$$\Pr[X \ge (j_2 - j_1) \cdot (2/10 + 2/10 - 0.9) + 3r] \le 3 \cdot e^{-\Omega(r)} + 1/ \poly(n),$$
which implies the lemma.
\end{proof}

Combining the previous lemmas, we can now obtain a bound on $j_2$ (which, in the following lemma is simply denoted by $j$).
\begin{lemma}
    Let $x$ be an element (maybe in the hash table) and let $h \in \{h_1(x), h_2(x)\}$. Let $j$ be the smallest position $j \ge h$ to contain an element $z$ using its first hash, and with first hash $> h$. Then,
    $$\E[j - h] = O(1).$$
    \label{lem:j2}
\end{lemma}
\begin{proof}
By Lemma \ref{lem:k123}, we have
$$j \le \max_{j_1\le h, j_2 \ge h} h + A_0 + A(h) - k_1(j_1, j_2) + 2k_2(j_1, j_2) + 2k_3(j_1, j_2),$$
where $A_0, k_1(j_1, j_2), k_2(j_1, j_2), k_3(j_1, j_2)$ are as defined in Lemma \ref{lem:k123}. It follows that
\begin{align*}
j_2 - h & \le A_0 +A(h) + \max_{j_1\le h, j_2 \ge h}  \max\left(0, -k_1(j_1, j_2) + 2k_2(j_1, j_2) + 2k_3(j_1, j_2)\right).
\end{align*}
The expected value of this is at most
\begin{align*}
    & \E[A(h)] + \E[A_0] + \sum_{j_1\le h, j_2 \ge h} \E\left[ \max\left(0, - k_1(j_1, j_2) + 2k_2(j_1, j_2) + 2k_3(j_1, j_2)\right)\right].
\end{align*}
By Lemma \ref{lem:Ah}, $\E[A(h)] = O(1)$, and the standard analysis of ordered linear probing, $\E[A_0] = O(1)$ (see, e.g., Corollary 2 of \cite{bender2022linear}). Finally, by Lemma \ref{lem:maxk123}, $\E\left[ \max\left(0, - k_1(j_1, j_2) + 2k_2(j_1, j_2) + 2k_3(j_1, j_2)\right)\right] \le 1 / 2^{\Omega(j_2 - j_1)} + 1/ \poly(n)$. Thus the overall expectation is at most
\begin{align*}
    & \le O(1) + \sum_{j_1\le h, j_2 \ge h} e^{-\Omega(j_2 - j_1)} \\
    & = O(1).
\end{align*}
\end{proof}

We now have all the pieces necessary to obtain time bounds for our algorithm.
\thmEvictions*
\begin{proof}
Consider an element $x$, and let $h \in \{h_1(x), h_2(x)\}$. By Lemma \ref{lem:j2}, the expected number of probes to positions $h, h + 1, h + 2, \ldots$ needed to find an element $y$ that is using its first hash and with $h_1(y) > h$ is $O(1)$. Since this holds for both $h = h_1(x)$ and $h_2(x)$, this implies an expected query time of $O(1)$.

For each insertion in the first phase, the expected insertion time is $O(1)$. Consider an insertion in the second phase, and suppose the hash table contains $(1 - \epsilon)n$ elements prior to the insertion. Then, each dart throw has probability $O(\sqrt{\epsilon})$ of success, so the expected number of dart throws that the insertion involves is $O(\sqrt{\epsilon}^{-1})$. Moreover, each dart throw involving an element $x$ takes time at most $O(|T(x)| + j(x) - h_2(x))$, where $|T(x)|$ is the size of the target zone containing $h_2(x)$ (or $0$ if $h_2(x)$ is not in a target zone), and where $j(x)$ is the first position at or after $h_2(x)$ to contain an element with hash $\ge h_2(x)$. By Lemmas \ref{lem:targetzone} and \ref{lem:j2}, $\E[|T(x)| + j_2(x) - h_2(x)] = O(1)$. Thus each dart throw takes expected time $O(1)$.

Let $X(i)$ be the event that the insertion performs at least $i$ dart throws, and let $R_i$ be the time spent on the $i$-th dart throw (or $0$ if no such dart throw exists). Then, the total expected insertion time is
$$O\left(\sum_{i = 1}^n \Pr[X(i)] \cdot \E[R_i \mid X(i)] \right).$$
Since each dart throw takes $O(1)$ expected time, this is
$$O(1) \cdot \sum_{i = 1}^n \Pr[X(i)].$$
Since $\sum_i \Pr[X(i)]$ is the expected number of dart throws, the sum evaluates to $O(\sqrt{\epsilon}^{-1})$, as desired.
\end{proof}

\bibliographystyle{alpha}
\bibliography{refs}

\end{document}